\documentclass[12pt,preprint]{aastex}

\usepackage{natbib}

\shorttitle{Near-Infrared Survey of EGOs, II}

\shortauthors{H.-T. Lee et al.}

\begin{document}



\title{Near-infrared H$_{2}$ and Continuum Survey of Extended Green Objects. II. Complete Census for the Northern Galactic Plane}

\author{Hsu-Tai Lee\altaffilmark{1} \email{htlee@illinois.edu}}
\affil{Institute of Astronomy and Astrophysics, Academia Sinica, P.O. Box 23-141, Taipei 106, Taiwan}
\altaffiltext{1}{Currently at University of Illinois at Urbana-Champaign}

\author{Wei-Ting Liao}
\affil{Department of Mechanical Engineering, National Taiwan University, Taiwan}

\author{Dirk Froebrich}
\affil{Centre for Astrophysics and Planetary Science, University of Kent, Canterbury, CT2 7NH, U.K.}

\author{Jennifer Karr}
\affil{Institute of Astronomy and Astrophysics, Academia Sinica, P.O. Box 23-141, Taipei 106, Taiwan}

\author{Georgios Ioannidis}
\affil{Centre for Astrophysics and Planetary Science, University of Kent, Canterbury, CT2 7NH, U.K.}

\author{Yong-Hyun Lee}
\affil{Department of Physics and Astronomy, Seoul National University, Seoul 151-742, Republic of Korea}

\author{Yu-Nung Su}
\affil{Institute of Astronomy and Astrophysics, Academia Sinica, P.O. Box 23-141, Taipei 106, Taiwan}

\author{Sheng-Yuan Liu}
\affil{Institute of Astronomy and Astrophysics, Academia Sinica, P.O. Box 23-141, Taipei 106, Taiwan}

\author{Hao-Yuan Duan}
\affil{Institute of Astronomy and Department of Physics, National Tsing-Hua University, Hsinchu 30013, Taiwan}

\author{Michihiro Takami}
\affil{Institute of Astronomy and Astrophysics, Academia Sinica, P.O. Box 23-141, Taipei 106, Taiwan}







\begin{abstract}

We discuss 94 Extended Green Objects (EGOs) in the northern Galactic plane cataloged by Cyganowski et al, based on near-infrared narrowband H$_{2}$ (2.122~$\micron$) and continuum observations from the United Kingdom Infrared Telescope.  This data set is three times larger than our previous study, and is unbiased by preselection.  As discussed in the previous paper, the morphologies of the 4.5~$\micron$ emission generally resemble those of the near-infrared continuum, but are different from those of the H$_{2}$ emission.  Of our sample, only 28\% of EGOs with H$_{2}$ emission show similar morphologies between 4.5~$\micron$ and H$_{2}$ emission.  These results suggest that the 4.5~$\micron$ emission mainly comes from scattered continuum from the embedded young stellar object (YSO), and partially from H$_{2}$ emission.  About half of EGOs are associated with H$_{2}$ outflows, if the H$_{2}$ outflow incompleteness is considered.  The H$_{2}$ outflow detection rate for EGOs with $K$-band detections (61\%) is significantly higher than for those without $K$-band detections (36\%).  This difference may be due to the fact that both H$_{2}$ and $K$-band emissions are associated with outflows, i.e., H$_{2}$ emission and $K$-band continuum are associated with shocks and outflow cavities, respectively.  We also compared the correlation between the H$_{2}$ outflows and Class I 44~GHz methanol masers from literature.  The methanol masers can be located upstream or downstream of the H$_{2}$ outflows and some bright H$_{2}$ spots or outflows are not associated with methanol masers, suggesting that methanol masers and H$_{2}$ emission trace different excitation conditions.


\end{abstract}


\section{INTRODUCTION}

The current understanding of massive star formation is very limited in both theory and observation \citep{zin07}.  For the observational point of view, they are rare, distant, and embedded.  The {\it Spitzer Space Telescope} provides a good opportunity for studying massive star formation, since it has good sensitivity, resolution, and a wide field of view.  The Galactic Legacy Infrared Mid-Plane Survey Extraordinaire \citep[GLIMPSE;][]{ben03} covers the inner Galactic plane where the majority of young stellar objects (YSOs) in the Milky Way are located.  The resolution at IRAC bands ($\sim2\arcsec$) can resolve the diffuse emission related to YSOs.  Based on this survey, \citet{cyg08} identified a new category of source, the ``Extended Green Object (EGO)''.  In the $Spitzer$ images (3.6, 4.5, and 8~$\micron$ bands in blue, green, and red, respectively), EGOs show extended emission in the green channel which could be due to excesses in the 4.5~$\micron$ band.  It is suggested that this emission may arise from H$_{2}$ ($\upsilon=0-0, S(9, 10, 11)$) and/or CO ($\upsilon=1-0$), excited by shocks from outflows \citep{cyg08,cha09}.  Therefore, EGOs are candidates for outflows from massive YSOs (MYSOs).  Follow-up studies implied that EGOs may be related to outflow activities \citep{cyg11,cyg09,che09,he12}.

Recently, \citet{tak12} used IRAC color-color diagrams to study the emission mechanism of EGOs.  They compared six EGOs with the colors of other objects and models of the scattered continuum, shocked H$_{2}$ emission, fluorescent H$_{2}$ emission, and PAH emission.  Of these emission mechanisms, the scattered continuum gives the simplest explanation for the observed results.  \citet{sim12} used the Infrared Spectrograph (IRS) on board $Spitzer$ to study several EGOs, with spectra covering 5--36~$\micron$.  They claimed that the 4.5~$\micron$ emission is not due to the H$_{2}$ lines, which are too faint to contribute to the 4.5~$\micron$ broad-band emission.

We have studied 34 EGOs \citep[][hereafter Paper I]{lee12}, using H$_{2}$ narrowband and near-infrared continuum observations from the United Kingdom Infrared Telescope (UKIRT).  Combined with $Spitzer$ IRAC, and MIPS images, we found that the H$_{2}$ features of the EGOs are more extended than the the 4.5~$\micron$ emission and $K$-/$H$-band continuum, and the H$_{2}$ emission is rarely associated with the 4.5~$\micron$ emission (3/34).  In addition, the morphologies of the near-infrared continuum resemble those of the 4.5~$\micron$ emission, suggesting that the 4.5~$\micron$ emission of the EGOs is dominated by scattered continuum from the embedded YSOs.

Some of the H$_{2}$ observations in Paper I were preselected, based on previous $K$-band observations.  Continuing from Paper I, we include the remaining 60 EGOs in the northern inner Galactic plane for further studies in this paper.  This not only increases the EGO sample by a factor of three (94 in total including those in Paper I), but also allows us to preform unbiased statistical studies of the EGOs.  Additionally, we discuss the relationship between Class I methanol masers and H$_{2}$ outflows in this paper.



This paper is organized as follows.  We describe the observations and results for the 60 EGOs in Section 2 and 3, respectively.  Together with the 34 EGOs in Paper~I, we discuss the near-infrared properties of the EGOs in the northern Galactic plane in Section 4, and summarize our conclusions in Section 5.

\section{OBSERVATIONS AND DATA REDUCTION}

Narrowband H$_{2}$ images were obtained through the UKIRT Wide Infrared Survey for H$_{2}$ \citep[UWISH2, $\upsilon$=1-0 $S$(1) at 2.122~$\micron$;][]{fro11a}.  UWISH2 completed an inner Galactic plane survey ($10\arcdeg \lesssim l \lesssim 65\arcdeg$; $-1\arcdeg.3 \lesssim b \lesssim +1\arcdeg.3$) with the Wide Field Camera (WFCAM) at UKIRT.  A narrowband filter ($\Delta\lambda=0.021~\micron$) at 2.122~$\micron$ was used to take H$_{2}$ images, with an integration time of 720~s per pixel.  The 5$\sigma$ detection limit magnitude of point sources is $\sim18$~mag in the H$_{2}$ narrow-band filter, and the surface brightness limit is $\sim10^{-19}$~W~m$^{-2}$ at the typical seeing \citep[$\sim0\arcsec.7$;][]{fro11a}.  Based on the estimations of \citet{tak10} for H$_{2}$ emission in the 4.5~$\micron$ band of IRAC, \citet{fro11a} suggested that the sensitivity of UWISH2 is 300-2000 times better than the corresponding GLIMPSE detections.  The UWISH2 data have been used to study the H$_{2}$ outflow activities for several star-forming regions \citep{fro11b,ioa12a,ioa12b,dew12}.

In the northern Galactic plane, there are 94 EGOs, which are cataloged by \citet{cyg08} and covered by UWISH2.  Paper~I presented the results of the H$_{2}$ and near-infrared continuum observations for 34 EGOs, and we show the remainder 60 EGOs here.

As for Paper I, we used $K$- and $H$-band images for studying the near-infrared continuum emission.  We accessed the $K$ (2.20~$\micron$) and $H$ (1.63~$\micron$)-band images through the WFCAM Science Archive for the UKIRT Infrared Deep Sky Survey Galactic Plane Survey \citep[UKIDSS GPS;][]{luc08}.  GPS surveys more than 1800 deg$^{2}$ along the northern equatorial Galactic plane in the near-infrared.  The 5$\sigma$ $H$ (1.63~$\micron$) and $K$ (2.20~$\micron$) detection limits for point sources are $\sim$19 and $\sim$18.5~mag, respectively.  A seeing of $<1\arcsec$ at the $K$-band is requested by the survey.  The $K$-band images were used to create the continuum-subtracted H$_{2}$ images, and to compare with the morphologies of the 4.5~$\micron$ emission.

The near-infrared images have been reduced by the Cambridge Astronomical Survey Unit (CASU).  The data reduction included flat fielding and sky correction, and frame stacking \citep{dye06}.  The reduced images were obtained via the Wide Field Astronomy Unit\footnote{http://www.roe.ac.uk/ifa/wfau/}.


The seeing conditions are different for the H$_{2}$ and $K$ observations.  Before subtracting the continuum, the H$_{2}$ and $K$-band images have to be smoothed to the same full width at half-maximum and the flux of the $K$-band image is scaled to the same flux as the H$_{2}$ image.  The continuum-subtracted H$_{2}$ image is generated by subtracting the smoothed and scaled $K$-band image from the corresponding H$_{2}$ narrowband image. For the bright stars with significant residuals in the difference image, we perform PSF-fitting photometry, and subtract a flux-scaled PSF sample image which is locally extracted from the bright reference stars around the target.  See the forthcoming paper (Lee J.-J. et al. 2013 in preparation) for a detailed description of the entire reduction procedure.

We visually inspected the continuum-subtracted H$_{2}$ images for the positions of the 60 EGOs.  Once we found an elongated H$_{2}$ outflow or H$_{2}$ knot, we checked its H$_{2}$ and K images to avoid contamination from image artifacts.

In addition to the near-infrared images, we obtained the $Spitzer$ IRAC and MIPS 24~$\micron$ images from the GLIMPSE \citep{chu09} and MIPSGAL \citep{car09}, respectively.  We downloaded the post-Basic Calibrated Data mosaic images from NASA/IPAC Infrared Science Archive (IRSA)\footnote{http://irsa.ipac.caltech.edu/}.  The IRAC images provide the morphologies of the EGOs, which will be compared to the H$_{2}$ and the near-infrared continuum emission.  The MIPS 24~$\micron$ images are mainly used to locate the positions of the YSOs in the EGOs.  In addition, we also downloaded the IRAC images for 6 low-mass protostellar outflows from \citet{tak10} for comparison.

\section{RESULTS}

\begin{figure}
\includegraphics[angle=0,scale=0.68]{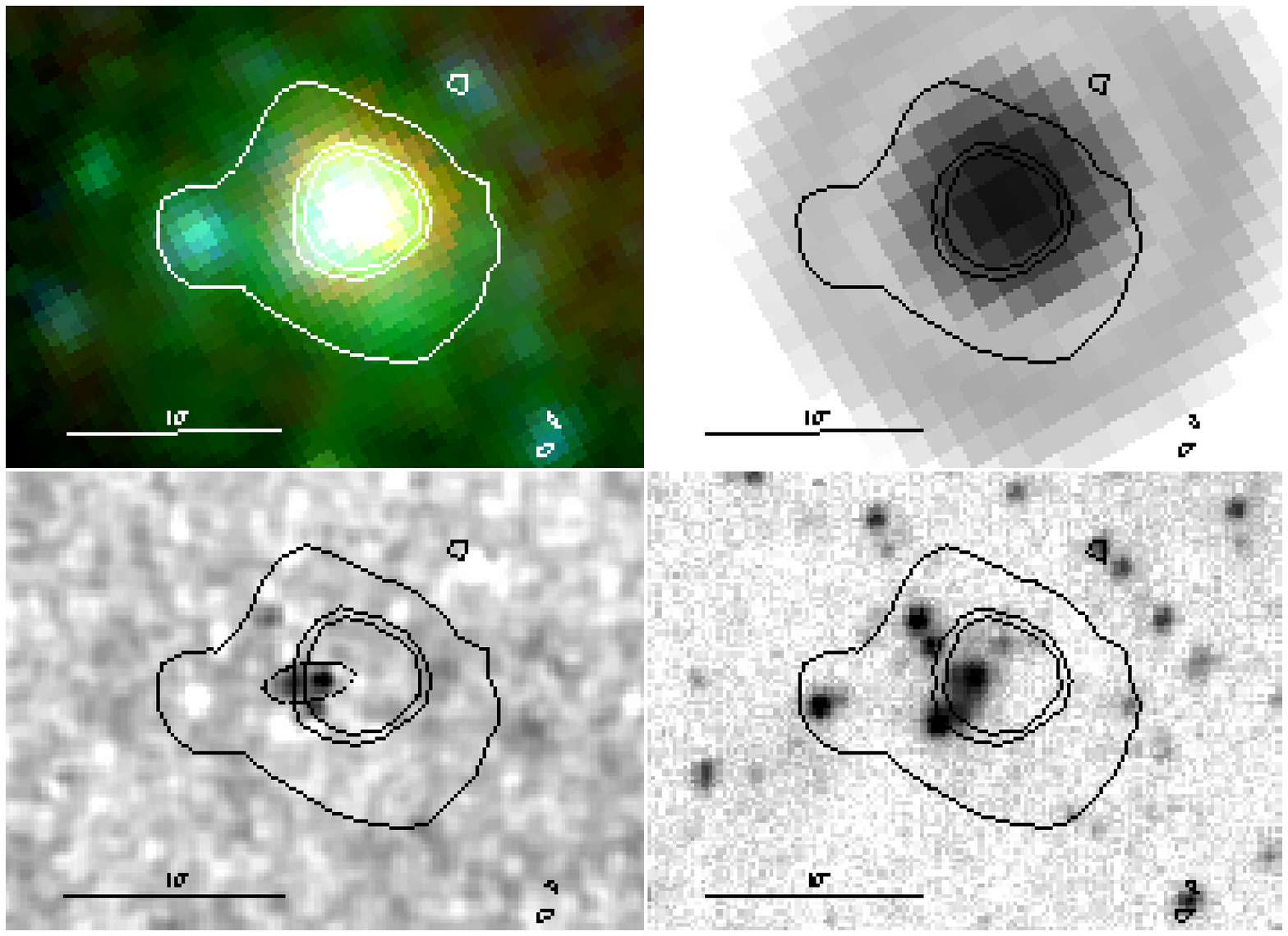}
\caption{Images of EGO G11.11-0.11.  Upper left panel: {\it Spitzer} IRAC image showing 3.6~$\micron$ (blue), 4.5~$\micron$ (green), and 8.0~$\micron$ (red); upper right panel: $Spitzer$ MIPS 24~$\micron$ image; lower left panel: continuum-subtracted H$_{2}$ image; lower right panel: GPS $K$-band image.  Darker regions represent higher flux values in the 24~$\micron$, continuum-subtracted H$_{2}$, and $K$-band images.  In addition, 4.5~$\micron$ contours are superposed.  The contour levels are chosen arbitrarily for the best comparisons of the flux distribution between the 4.5~$\micron$ emission and the others.  In the continuum-subtracted H$_{2}$ image, compact features with a combination of positive and negative valued features are the residual of continuum subtraction of point sources.  The resolutions of IRAC, MIPS, continuum-subtracted H$_{2}$, and $K$-band images are $\sim2\arcsec$, 6$\arcsec$, $\lesssim1\arcsec$, and $\sim1\arcsec$, respectively.  In the continuum subtracted H$_{2}$ image, the black dashed ellipse marks the H$_{2}$ lobe (MHO 2304).}
\label{fig:G11.11-0.11}
\end{figure}

\begin{figure}
\includegraphics[angle=0,scale=0.55]{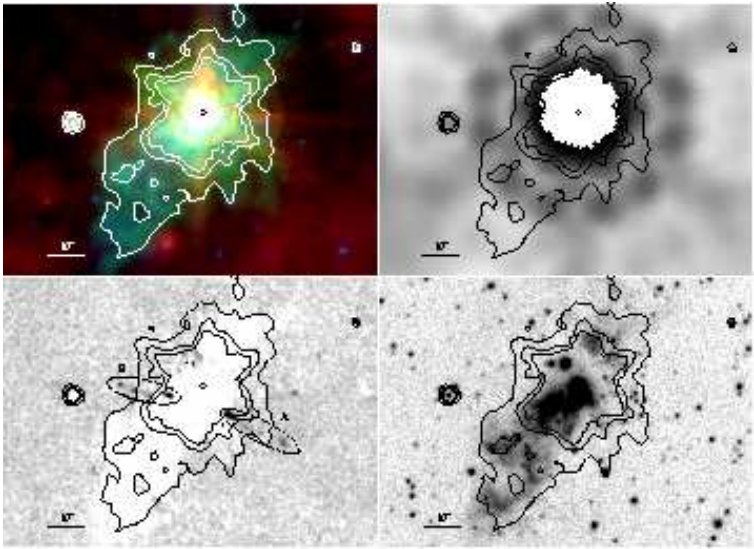}
\caption{Same as Figure~\ref{fig:G11.11-0.11}, but for EGO G12.91-0.26.  The position of W33A is labeled by black diamond.  The white part in the 24~$\micron$ source is saturated.  In the continuum-subtracted H$_{2}$ image, negative valued features (white) are likely to represent continuum emission with a large infrared excess and high extinction, and may be scattered continuum from the YSOs.  The black dashed ellipses illustrate a bipolar H$_{2}$ outflow (MHO 2305).}
\label{fig:G12.91-0.26}
\end{figure}

\begin{figure}
\includegraphics[angle=0,scale=0.55]{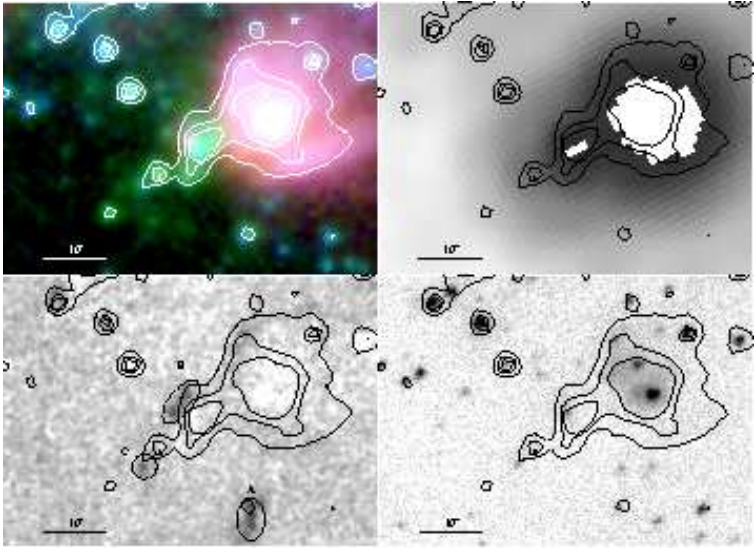}
\caption{Same as Figure~\ref{fig:G11.11-0.11}, but for EGO G14.33-0.64.  The white region in the 24~$\micron$ image is saturated.  The black dashed ellipses A and B illustrate two H$_{2}$ lobes from the EGO (MHO 2306).}
\label{fig:G14.33-0.64}
\end{figure}

\begin{figure}
\includegraphics[angle=0,scale=0.68]{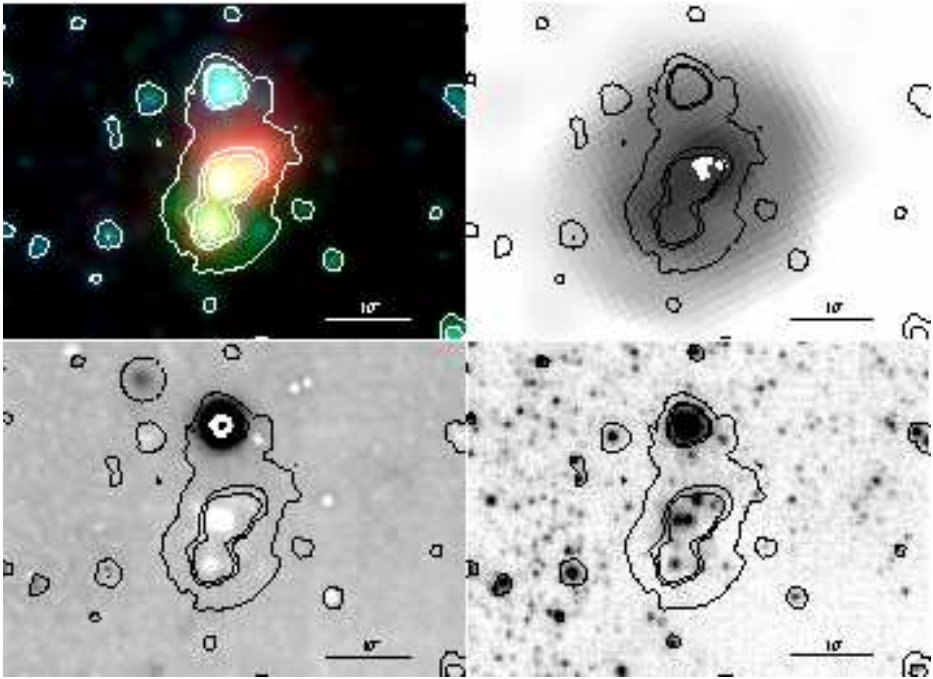}
\caption{Same as Figure~\ref{fig:G11.11-0.11}, but for EGO G17.96+0.08.}
\label{fig:G17.96+0.08}
\end{figure}

\begin{figure}
\includegraphics[angle=0,scale=0.68]{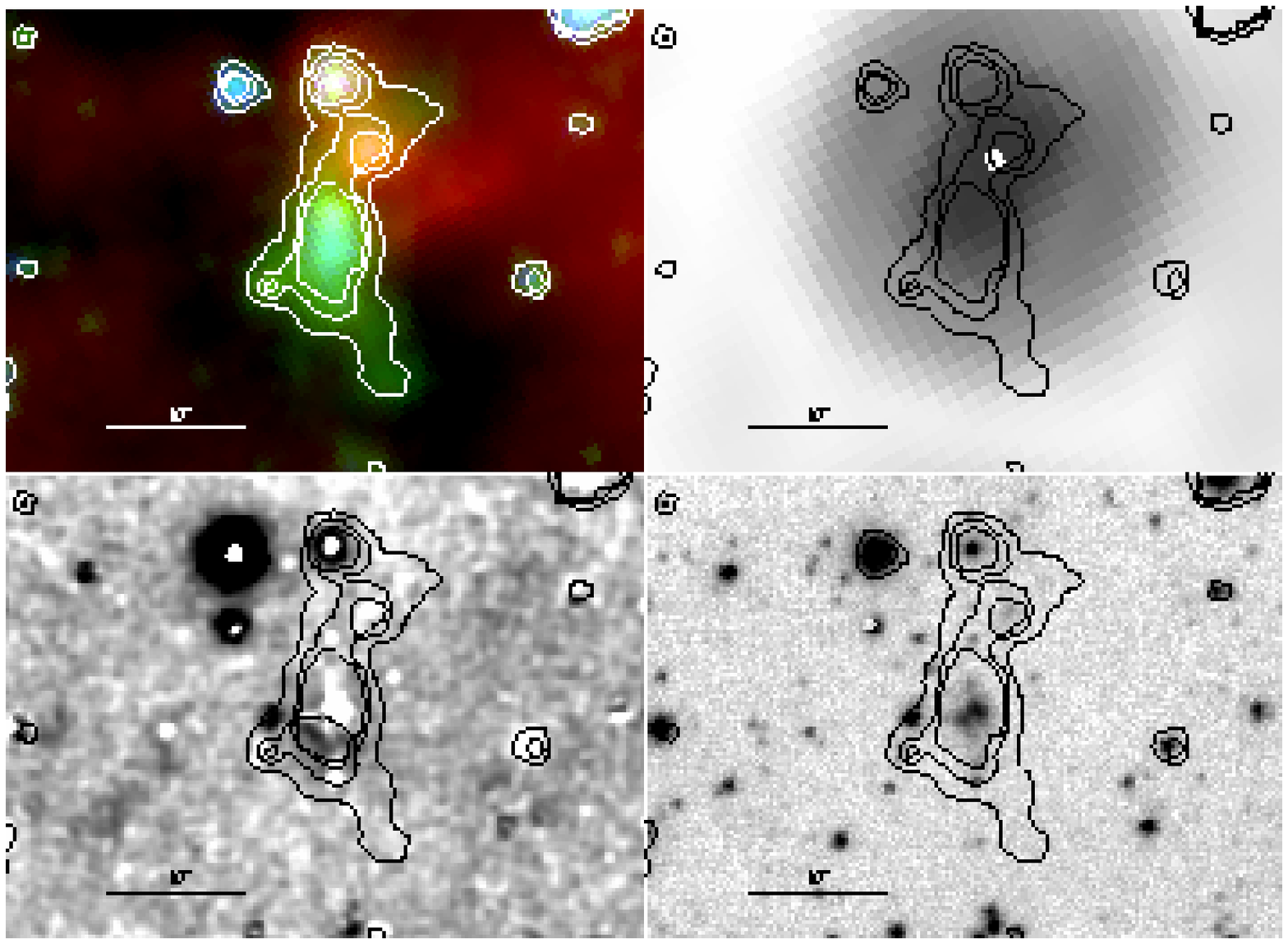}
\caption{Same as Figure~\ref{fig:G11.11-0.11}, but for EGO G19.36-0.03.}
\label{fig:G19.36-0.03}
\end{figure}

\begin{figure}
\includegraphics[angle=0,scale=0.68]{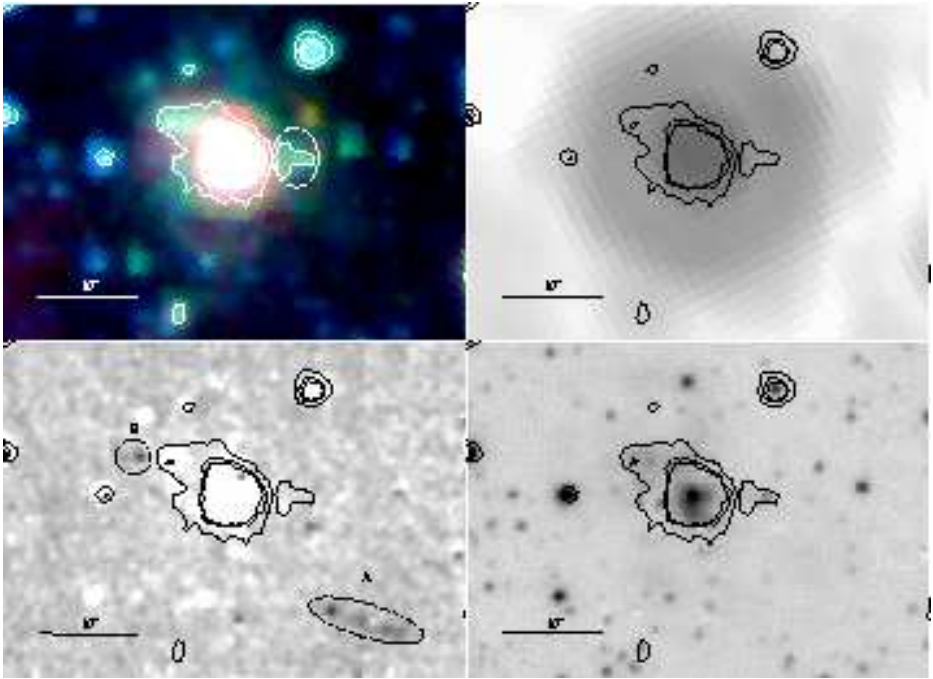}
\caption{Same as Figure~\ref{fig:G11.11-0.11}, but for EGO G21.24+0.19.  The position of EGO G21.24+0.19 is marked by a white dashed ellipse.}
\label{fig:G21.24+0.19}
\end{figure}

\begin{figure}
\includegraphics[angle=0,scale=0.68]{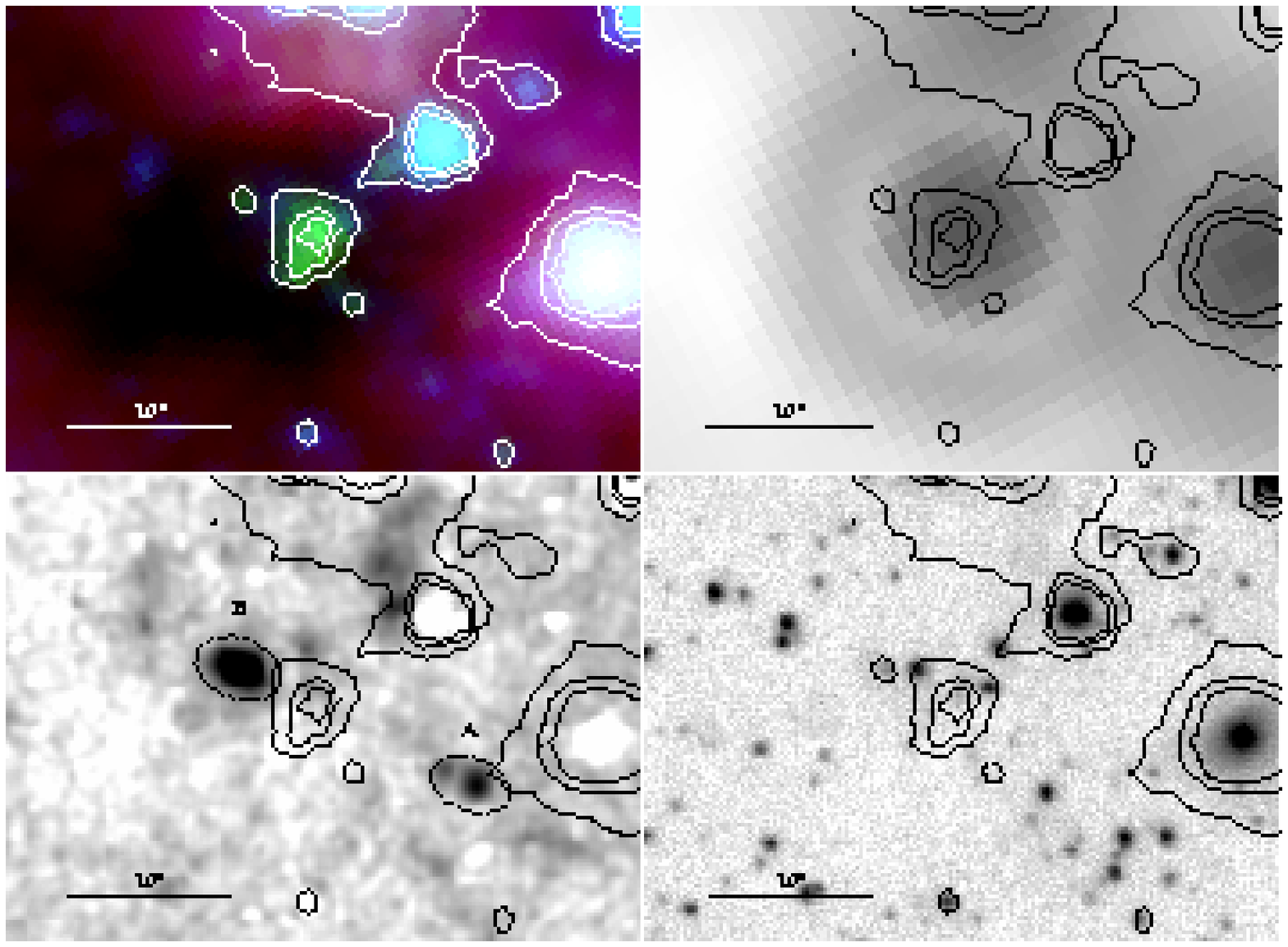}
\caption{Same as Figure~\ref{fig:G11.11-0.11}, but for EGO G22.04+0.22.}
\label{fig:G22.04+0.22}
\end{figure}

\begin{figure}
\includegraphics[angle=0,scale=0.65]{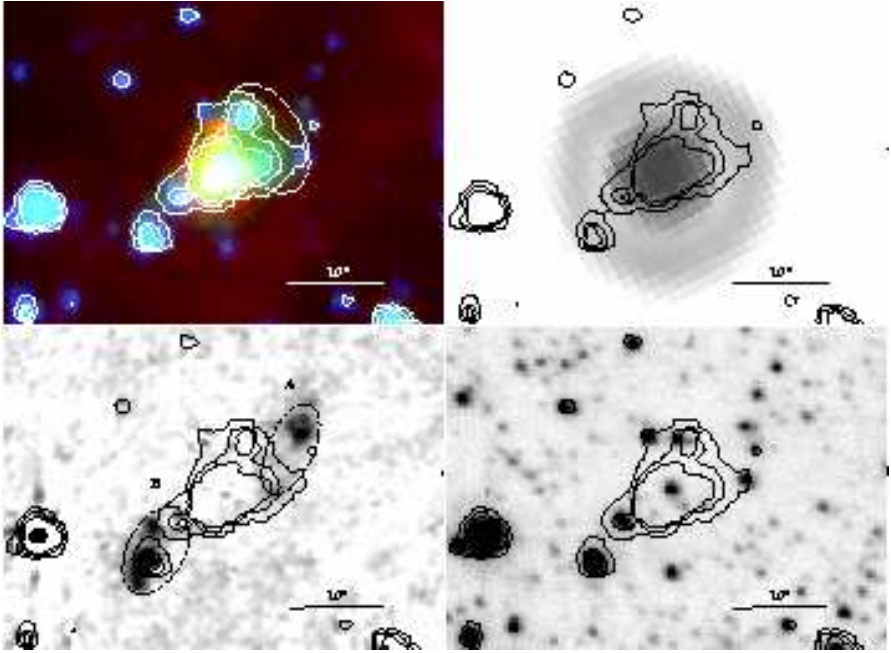}
\caption{Same as Figure~\ref{fig:G11.11-0.11}, but for EGO G23.82+0.38.  The position of EGO G23.82+0.38 is marked by a white dashed ellipse.}
\label{fig:G23.82+0.38}
\end{figure}

\begin{figure}
\includegraphics[angle=0,scale=0.65]{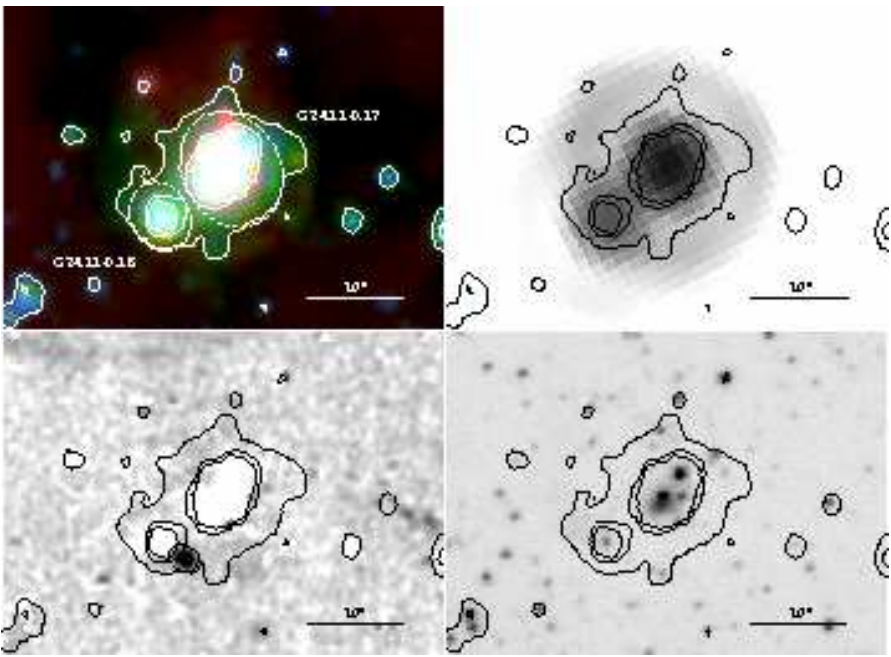}
\caption{Same as Figure~\ref{fig:G11.11-0.11}, but for EGO G24.11-0.17 and G24.11-0.18.  The positions of the two EGOs are marked by two white dashed ellipses.}
\label{fig:G24.11-0.17}
\end{figure}

\begin{figure}
\includegraphics[angle=0,scale=0.65]{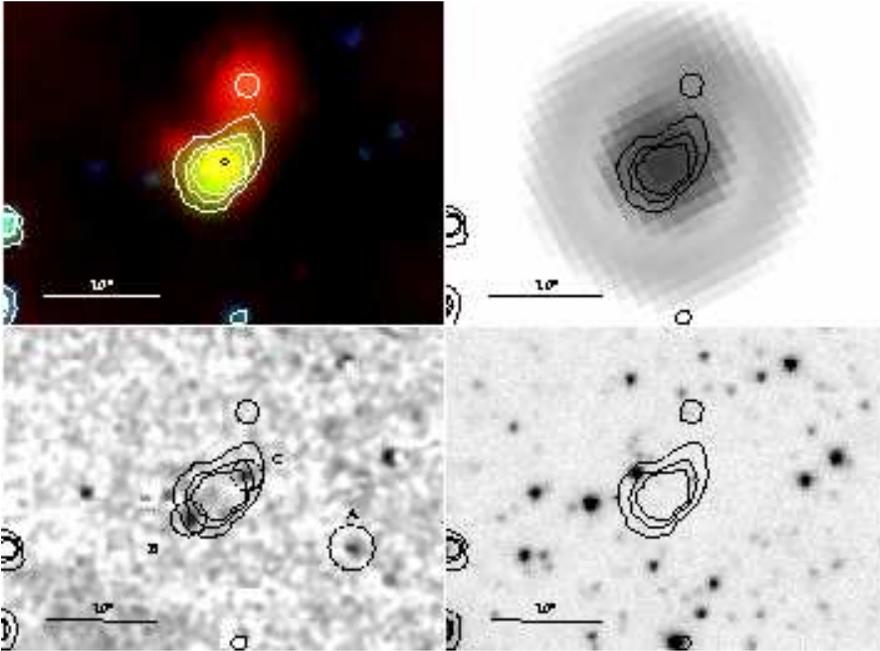}
\caption{Same as Figure~\ref{fig:G11.11-0.11}, but for EGO G24.33+0.14.  The position of an \ion{H}{2} region, G24.33+0.11 GLM1, is labeled by a black diamond.}
\label{fig:G24.33+0.14}
\end{figure}

\begin{figure}
\includegraphics[angle=0,scale=0.65]{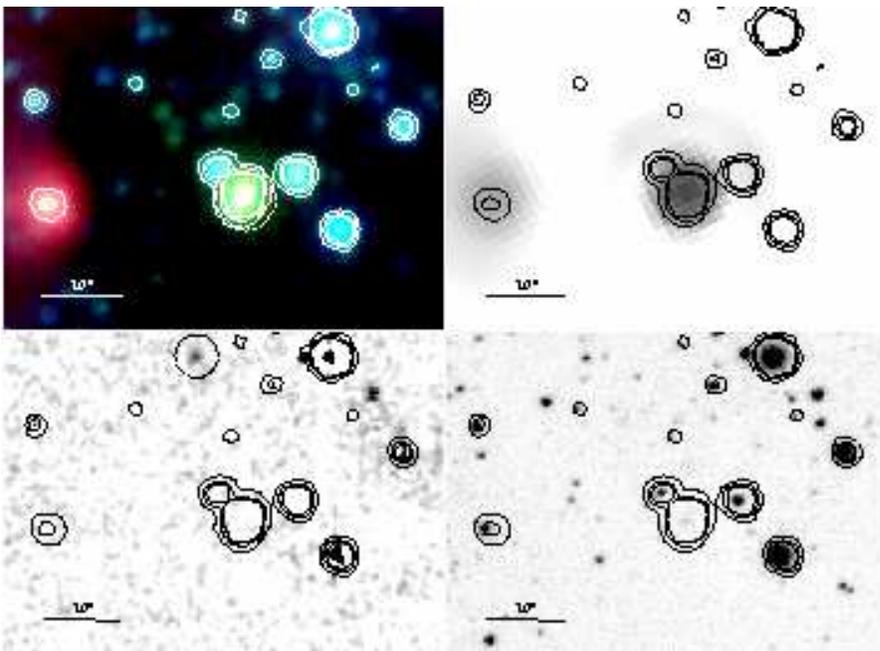}
\caption{Same as Figure~\ref{fig:G11.11-0.11}, but for EGO G24.63+0.15.  The position of EGO G24.63+0.15 is marked by a white dashed circle.}
\label{fig:G24.63+0.15}
\end{figure}

\begin{figure}
\includegraphics[angle=0,scale=0.65]{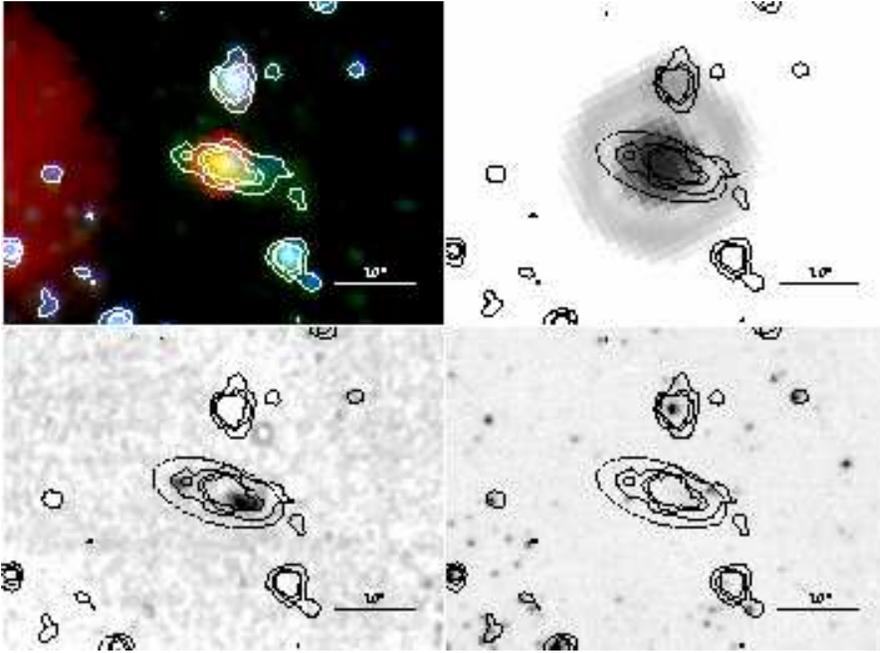}
\caption{Same as Figure~\ref{fig:G11.11-0.11}, but for EGO G27.97-0.47.}
\label{fig:G27.97-0.47}
\end{figure}

\begin{figure}
\includegraphics[angle=0,scale=0.65]{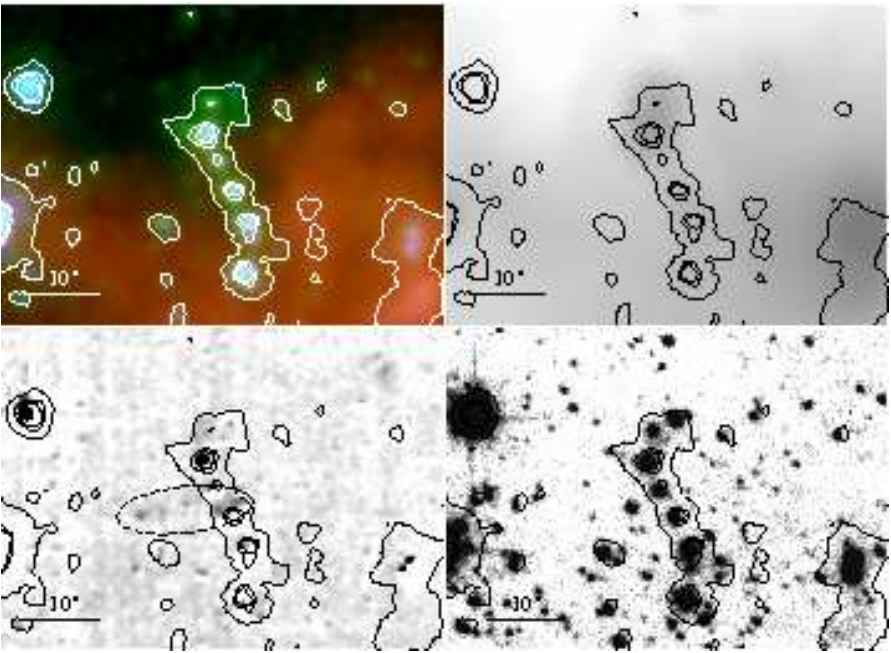}
\caption{Same as Figure~\ref{fig:G11.11-0.11}, but for EGO G29.96-0.79.  The position of EGO G29.96-0.79 is marked by a white dashed ellipse.  The black ellipse illustrates the H$_{2}$ outflow (MHO 2458).}
\label{fig:G29.96-0.79}
\end{figure}

\begin{figure}
\includegraphics[angle=0,scale=0.65]{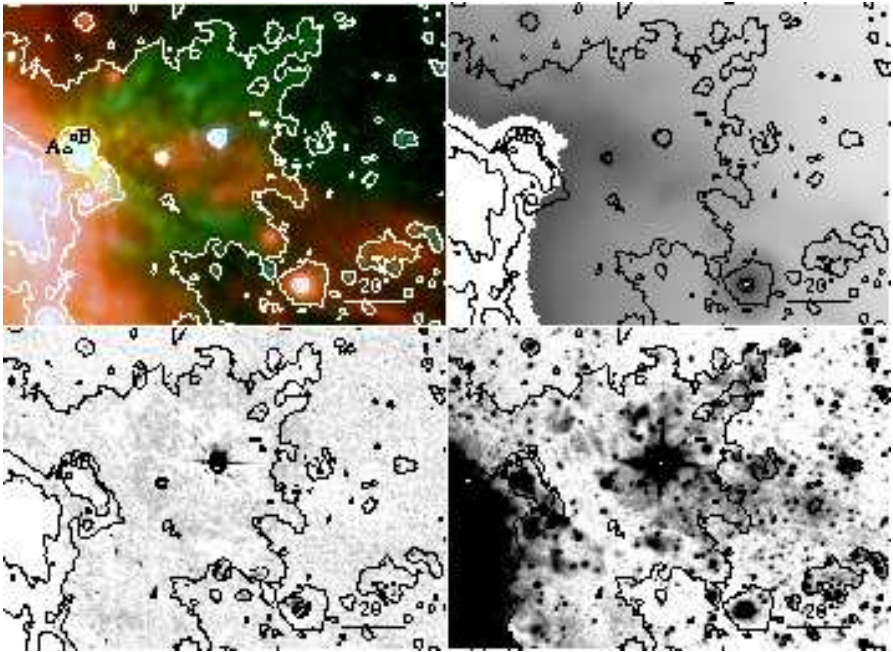}
\caption{Same as Figure~\ref{fig:G11.11-0.11}, but for EGO G34.26+0.15.  Two HC~\ion{H}{2} regions are labeled by circles (G34.26+0.15~A and B).  Three dashed eclipses indicate the positions of H$_{2}$ knots.}
\label{fig:G34.26+0.15}
\end{figure}

\begin{figure}
\includegraphics[angle=0,scale=0.65]{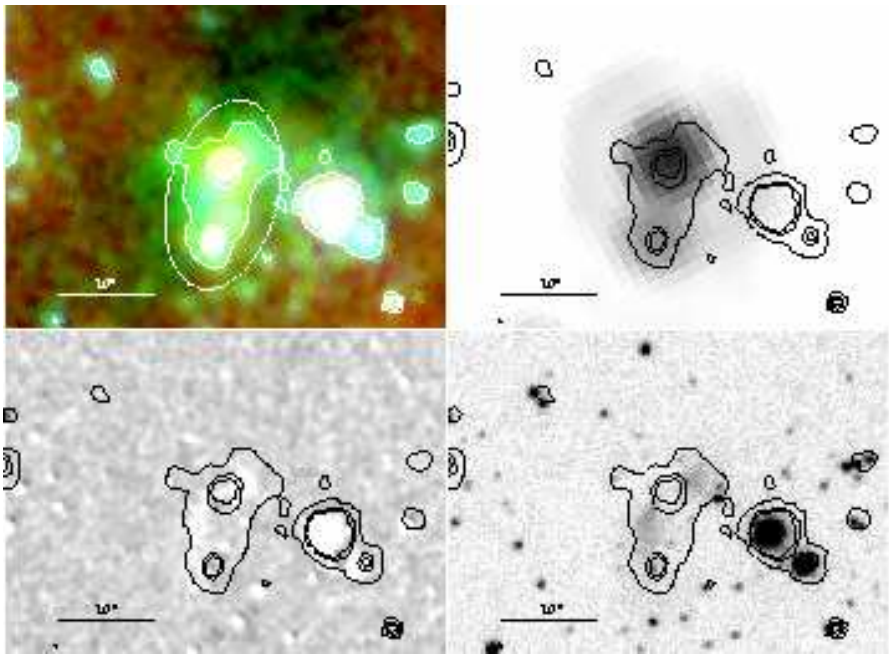}
\caption{Same as Figure~\ref{fig:G11.11-0.11}, but for EGO G34.28+0.18.  The position of EGO G34.28+0.18 is marked by a white dashed ellipse.}
\label{fig:G34.28+0.18}
\end{figure}

\begin{figure}
\includegraphics[angle=0,scale=0.65]{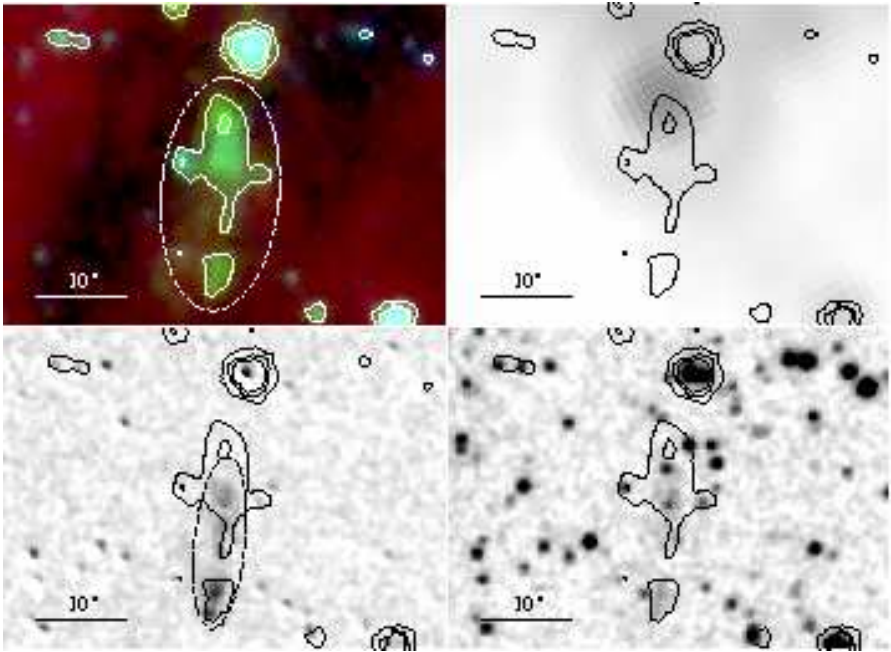}
\caption{Same as Figure~\ref{fig:G11.11-0.11}, but for EGO G34.39+0.22.  The position of EGO G34.39+0.22 is marked by a white dashed ellipse.  The black ellipse outlines the H$_{2}$ lobe (MHO 2459).}
\label{fig:G34.39+0.22}
\end{figure}

\begin{figure}
\includegraphics[angle=0,scale=0.65]{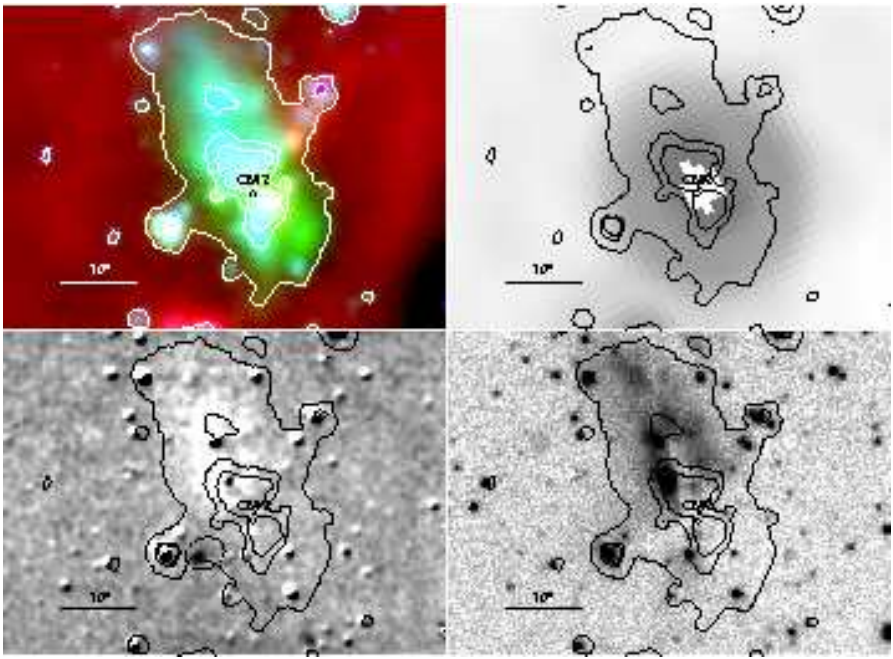}
\caption{Same as Figure~\ref{fig:G11.11-0.11}, but for EGO G35.03+0.35.  The position of EGO G35.03+0.35 CM2 is marked by a black diamond.  The black ellipse outlines the H$_{2}$ outflow (MHO 2460).}
\label{fig:G35.03+0.35}
\end{figure}

\begin{figure}
\includegraphics[angle=0,scale=0.65]{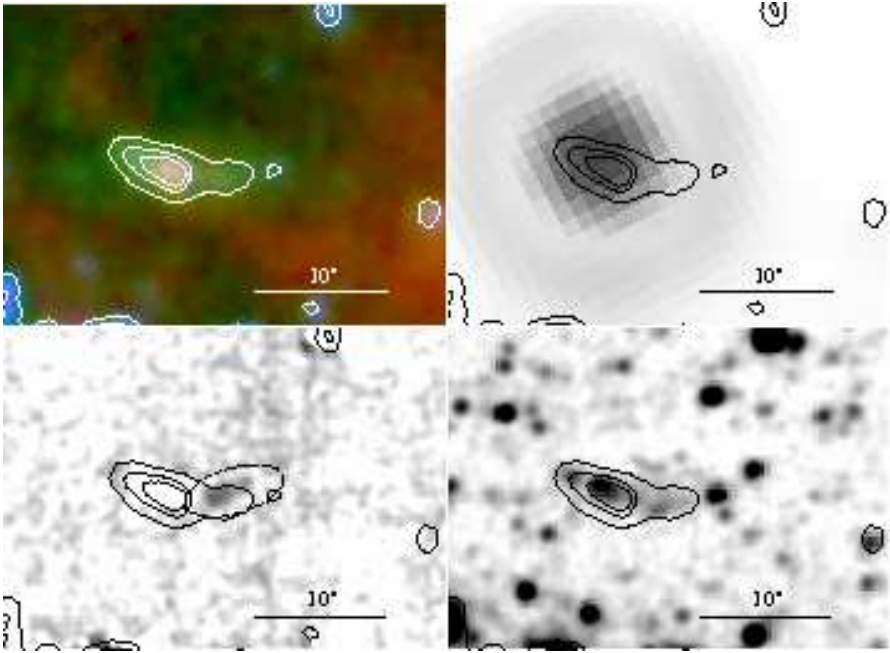}
\caption{Same as Figure~\ref{fig:G11.11-0.11}, but for EGO G37.48-0.10.  The black ellipse marks the H$_{2}$ lobe (MHO 2461).}
\label{fig:G37.48-0.10}
\end{figure}

\begin{figure}
\includegraphics[angle=0,scale=0.65]{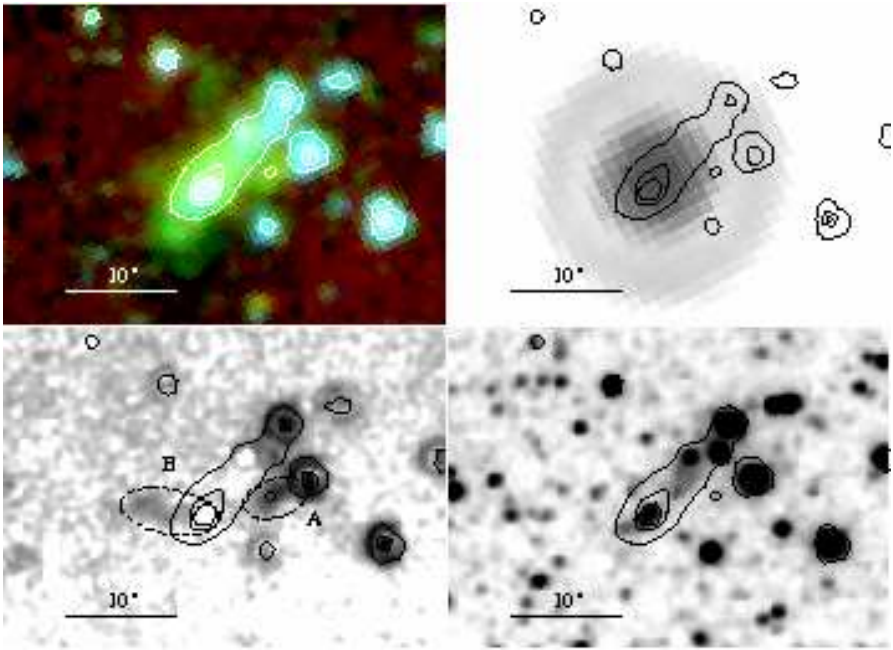}
\caption{Same as Figure~\ref{fig:G11.11-0.11}, but for EGO G39.10+0.49.  The black ellipses marks the H$_{2}$ outflow (MHO 2462).}
\label{fig:G39.10+0.49}
\end{figure}

\begin{figure}
\includegraphics[angle=0,scale=0.65]{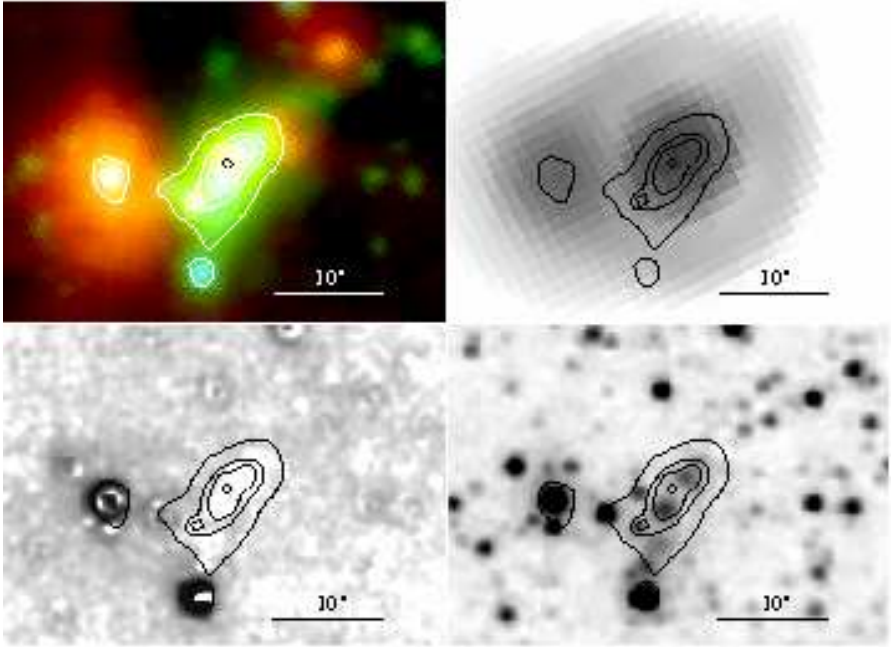}
\caption{Same as Figure~\ref{fig:G11.11-0.11}, but for EGO G39.39-0.14.  The diamond marks the position of the \ion{H}{2} region IRAS 19012+0536 B.}
\label{fig:G39.39-0.14}
\end{figure}

\begin{figure}
\includegraphics[angle=0,scale=0.65]{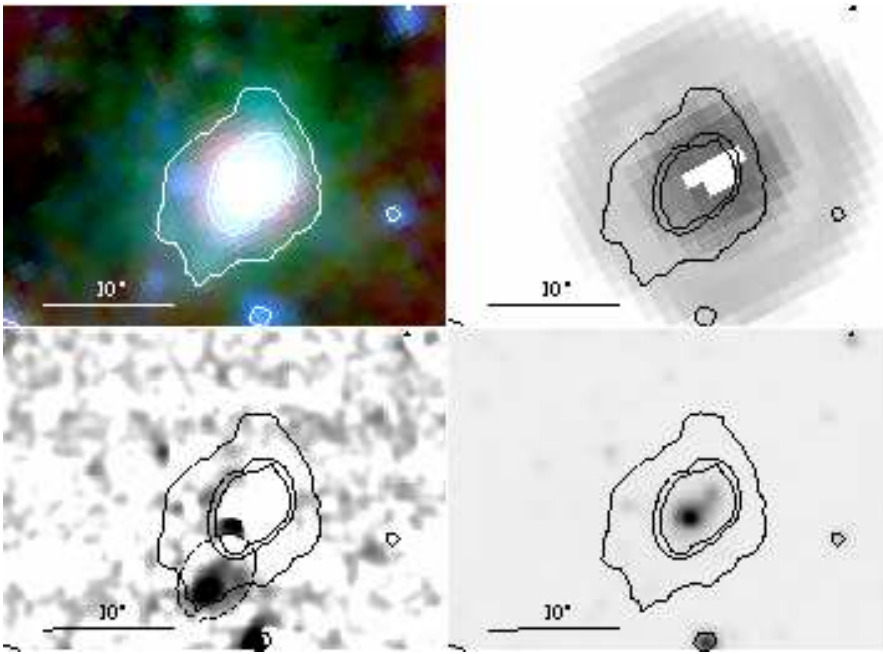}
\caption{Same as Figure~\ref{fig:G11.11-0.11}, but for EGO G40.28-0.22.  The black ellipse outlines the H$_{2}$ lobe (MHO 2463).}
\label{fig:G40.28-0.22}
\end{figure}

\begin{figure}
\includegraphics[angle=0,scale=0.65]{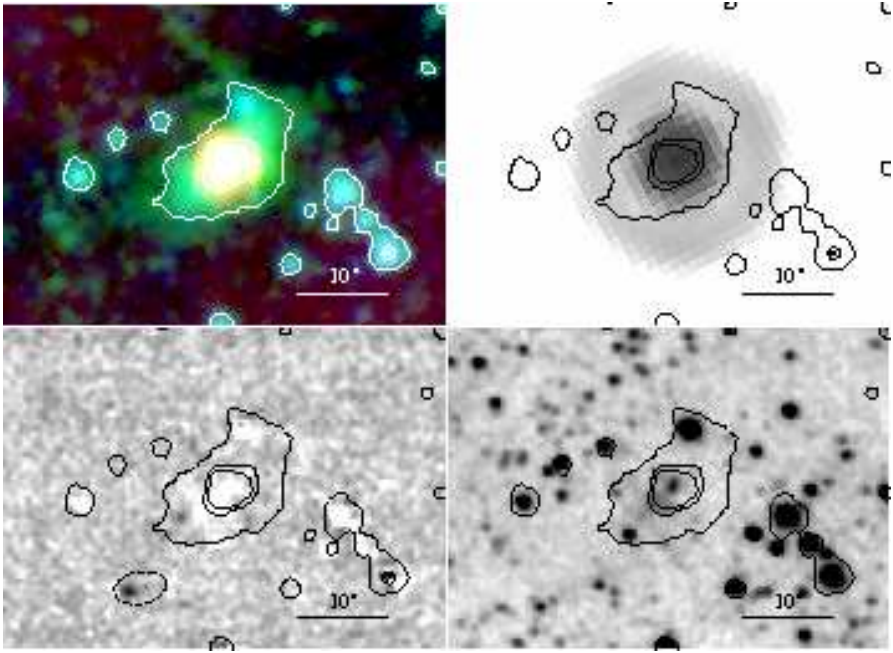}
\caption{Same as Figure~\ref{fig:G11.11-0.11}, but for EGO G40.28-0.27.}
\label{fig:G40.28-0.27}
\end{figure}

\begin{figure}
\includegraphics[angle=0,scale=0.65]{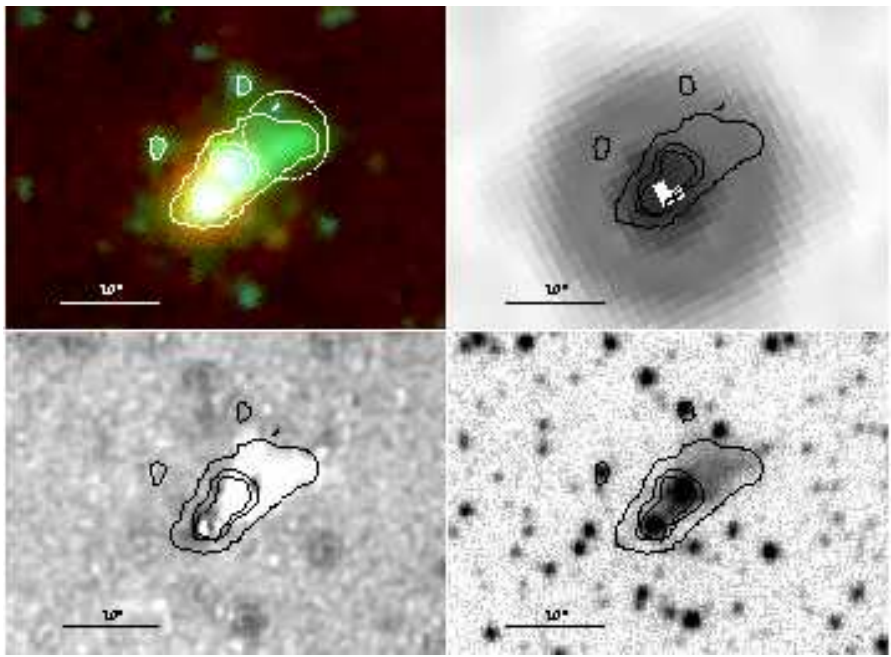}
\caption{Same as Figure~\ref{fig:G11.11-0.11}, but for EGO G40.60-0.72.  The position of EGO G40.60-0.72 is marked by a white dashed ellipse.}
\label{fig:G40.60-0.72}
\end{figure}

\begin{figure}
\includegraphics[angle=0,scale=0.65]{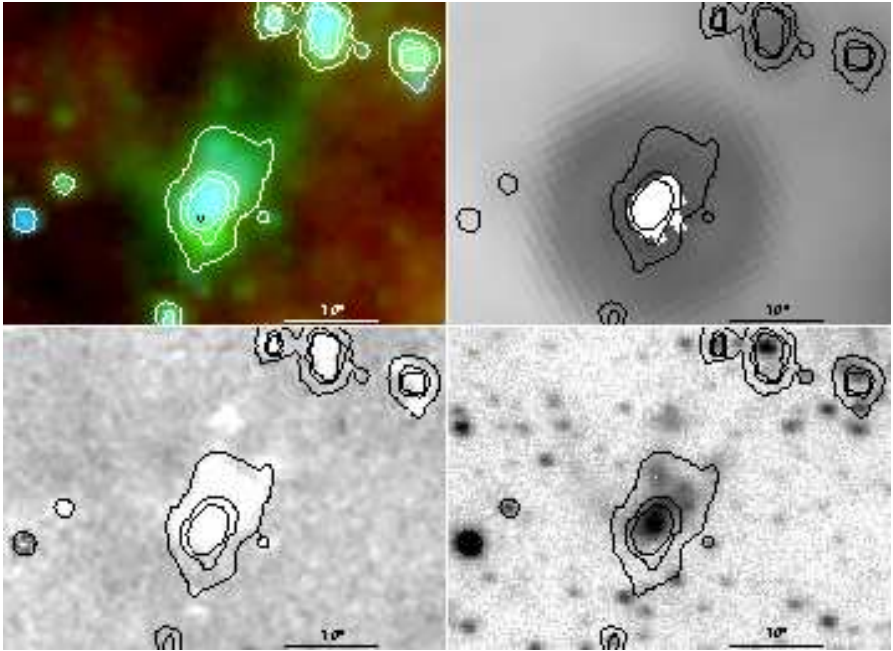}
\caption{Same as Figure~\ref{fig:G11.11-0.11}, but for EGO G45.47+0.05.  The position of the \ion{H}{2} region [UHP2009] VLA G045.4657+00.0453 is marked by a diamond.}
\label{fig:G45.47+0.05}
\end{figure}

\begin{figure}
\includegraphics[angle=0,scale=0.65]{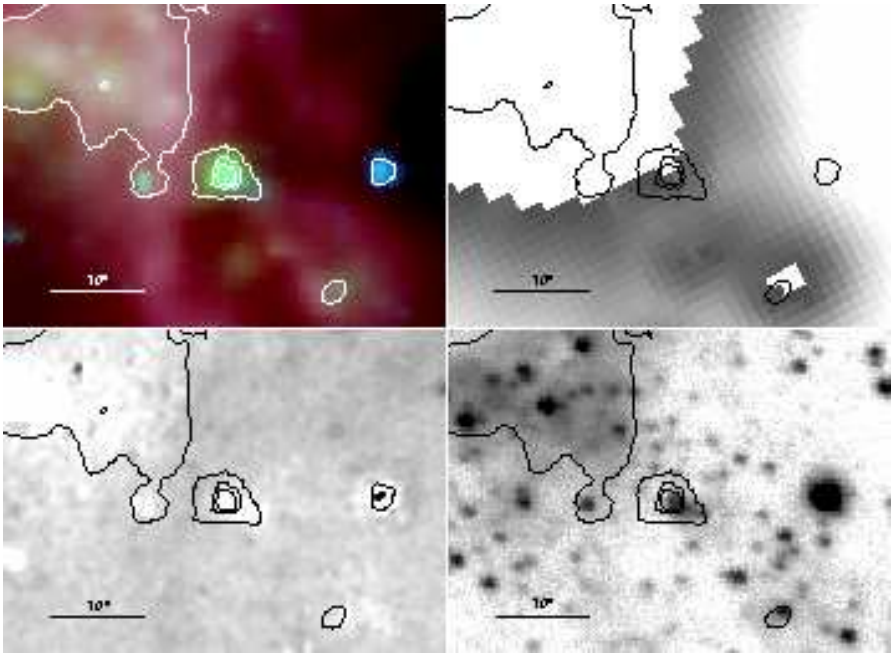}
\caption{Same as Figure~\ref{fig:G11.11-0.11}, but for EGO G45.47+0.13.}
\label{fig:G45.47+0.13}
\end{figure}

\begin{figure}
\includegraphics[angle=0,scale=0.65]{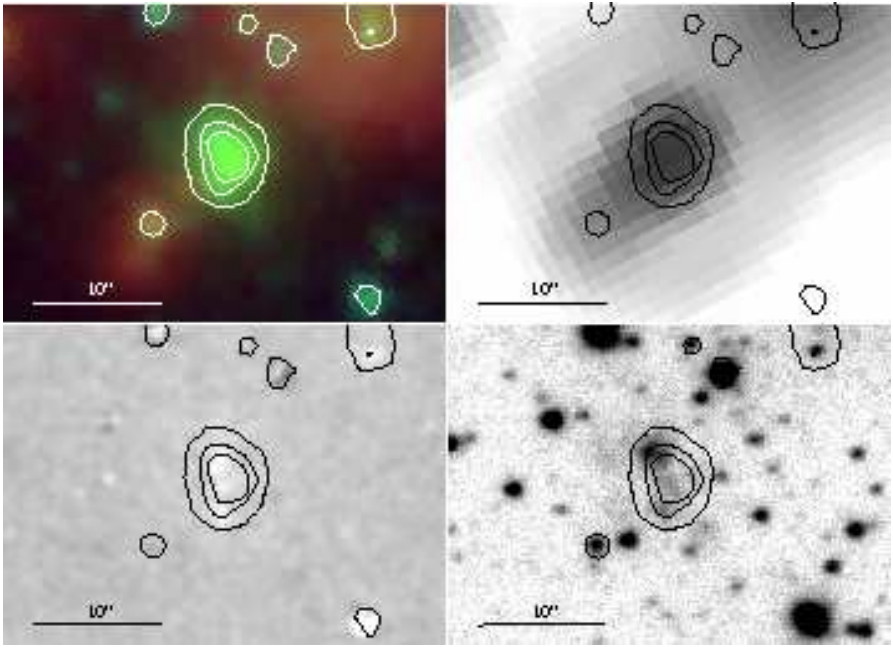}
\caption{Same as Figure~\ref{fig:G11.11-0.11}, but for EGO G45.80-0.36.}
\label{fig:G45.80-0.36}
\end{figure}

\begin{figure}
\includegraphics[angle=0,scale=0.65]{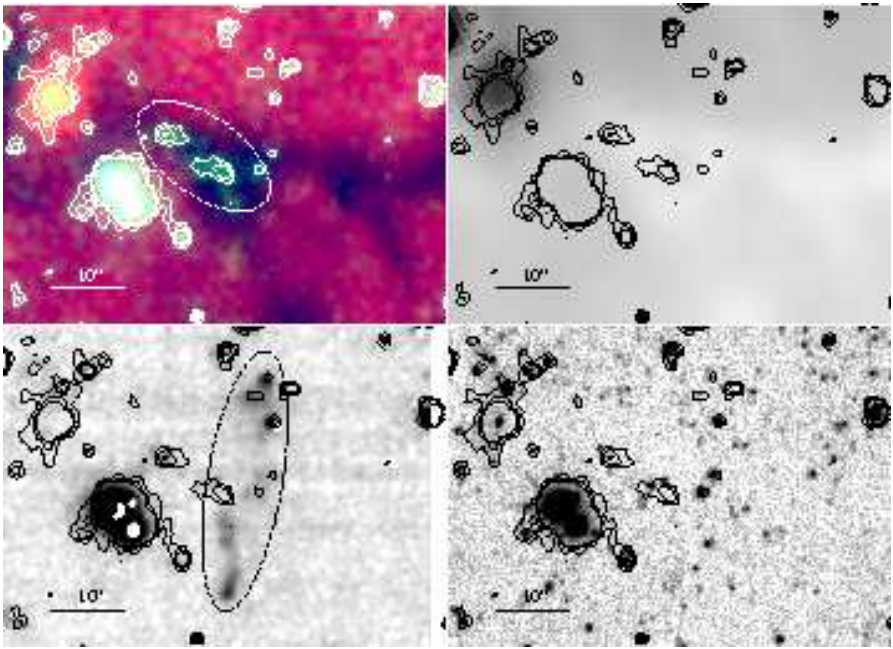}
\caption{Same as Figure~\ref{fig:G11.11-0.11}, but for EGO G48.66-0.30.  The position of EGO G48.66-0.30 is marked by a white dashed ellipse.  The black ellipse outlines the H$_{2}$ outflow (MHO 2465).}
\label{fig:G48.66-0.30}
\end{figure}

\begin{figure}
\includegraphics[angle=0,scale=0.65]{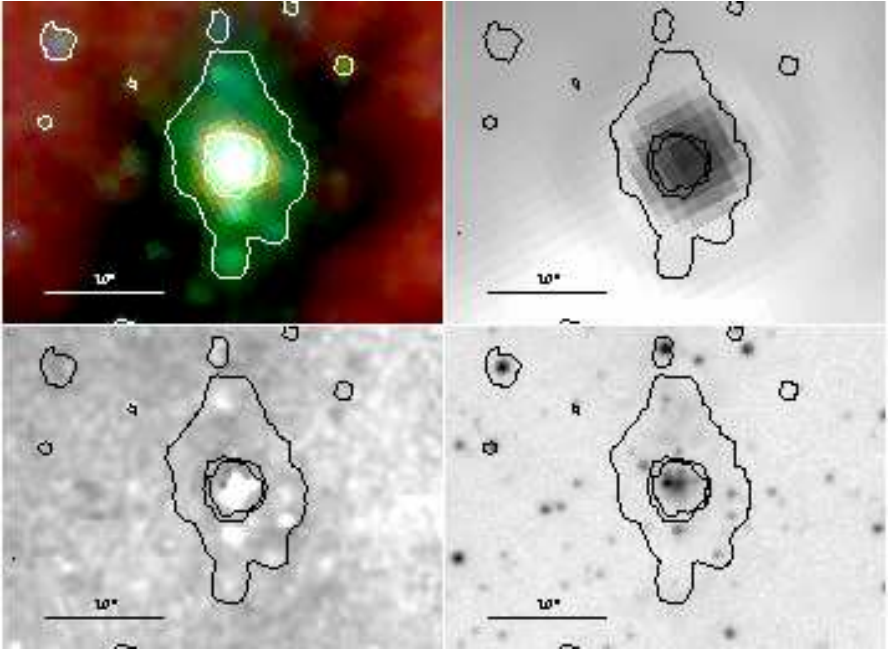}
\caption{Same as Figure~\ref{fig:G11.11-0.11}, but for EGO G49.07-0.33.}
\label{fig:G49.07-0.33}
\end{figure}

\begin{figure}
\includegraphics[angle=0,scale=0.65]{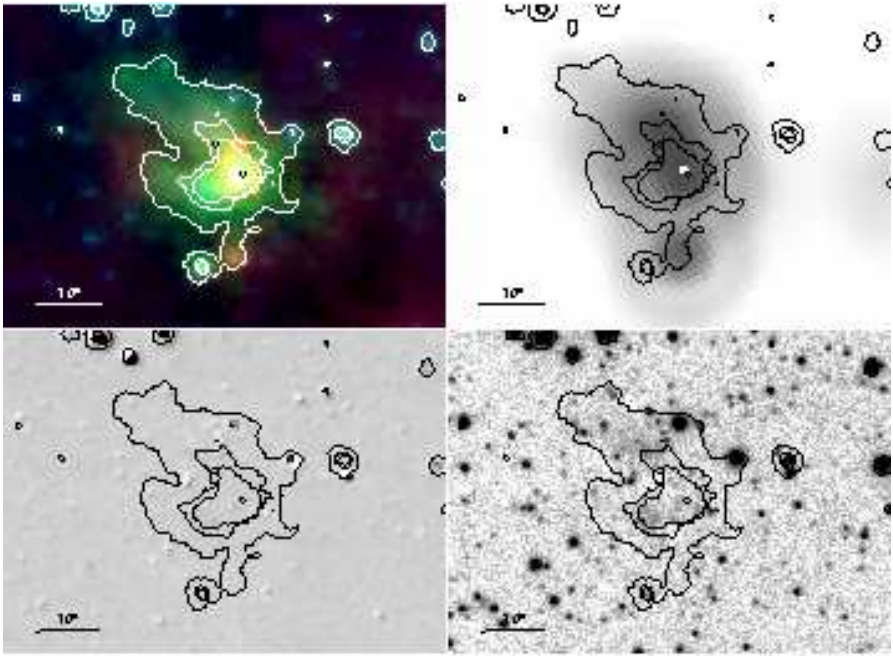}
\caption{Same as Figure~\ref{fig:G11.11-0.11}, but for EGO G49.27-0.34.  Two radio sources, reported by \citet{cyg11}, are labeled (diamonds).}
\label{fig:G49.27-0.34}
\end{figure}

\begin{figure}
\includegraphics[angle=0,scale=0.65]{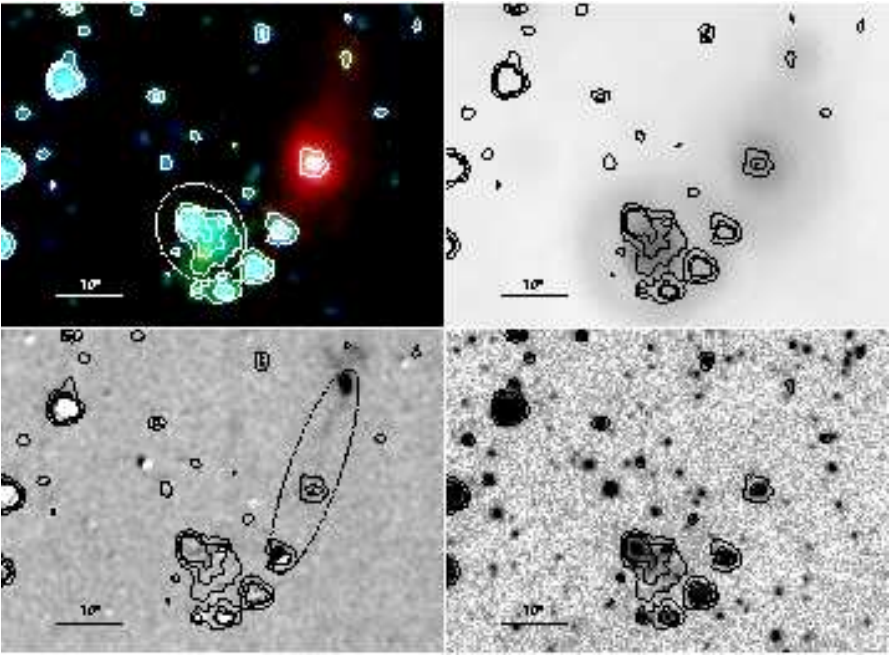}
\caption{Same as Figure~\ref{fig:G11.11-0.11}, but for EGO G50.36-0.42.  The position of EGO G50.36-0.42 is marked by a white dashed ellipse.  The black ellipse outlines the H$_{2}$ outflow (MHO 2624).}
\label{fig:G50.36-0.42}
\end{figure}

\begin{figure}
\includegraphics[angle=0,scale=0.65]{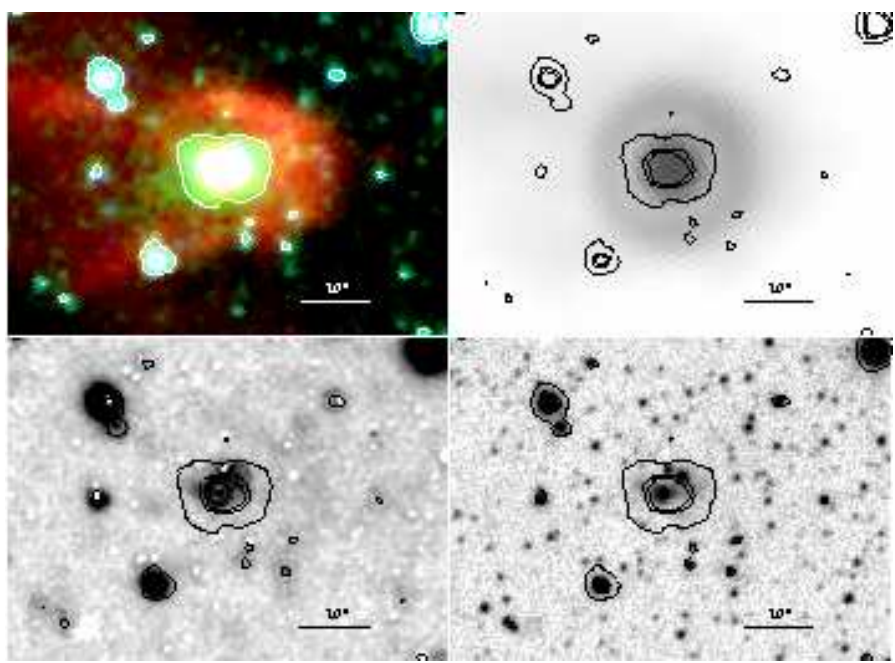}
\caption{Same as Figure~\ref{fig:G11.11-0.11}, but for EGO G53.92-0.07.}
\label{fig:G53.92-0.07}
\end{figure}

\begin{figure}
\includegraphics[angle=0,scale=0.65]{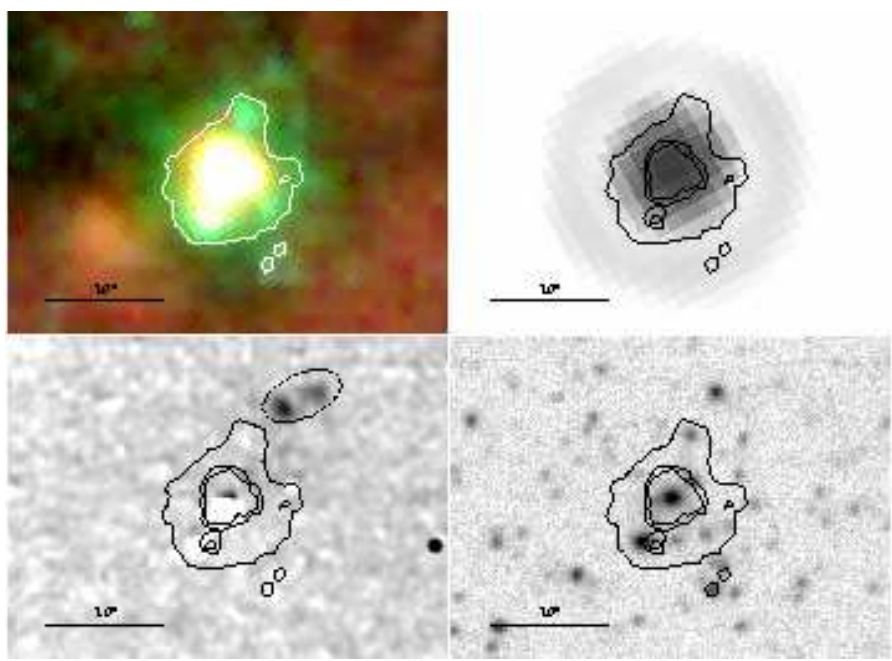}
\caption{Same as Figure~\ref{fig:G11.11-0.11}, but for EGO G54.11-0.08.}
\label{fig:G54.11-0.08}
\end{figure}

\begin{figure}
\includegraphics[angle=0,scale=0.65]{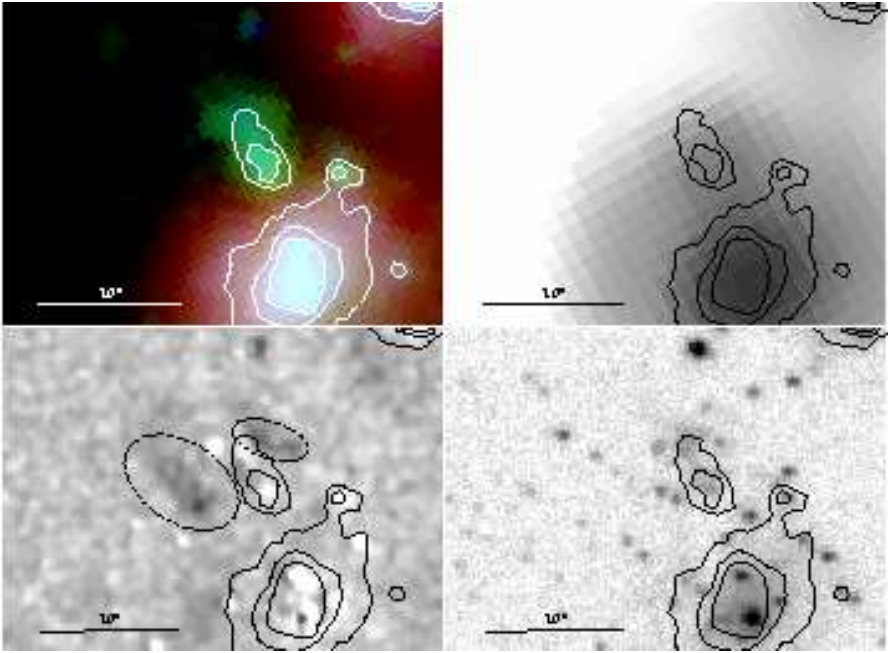}
\caption{Same as Figure~\ref{fig:G11.11-0.11}, but for EGO G57.61+0.02.}
\label{fig:G57.61+0.02}
\end{figure}

\begin{figure}
\includegraphics[angle=0,scale=0.65]{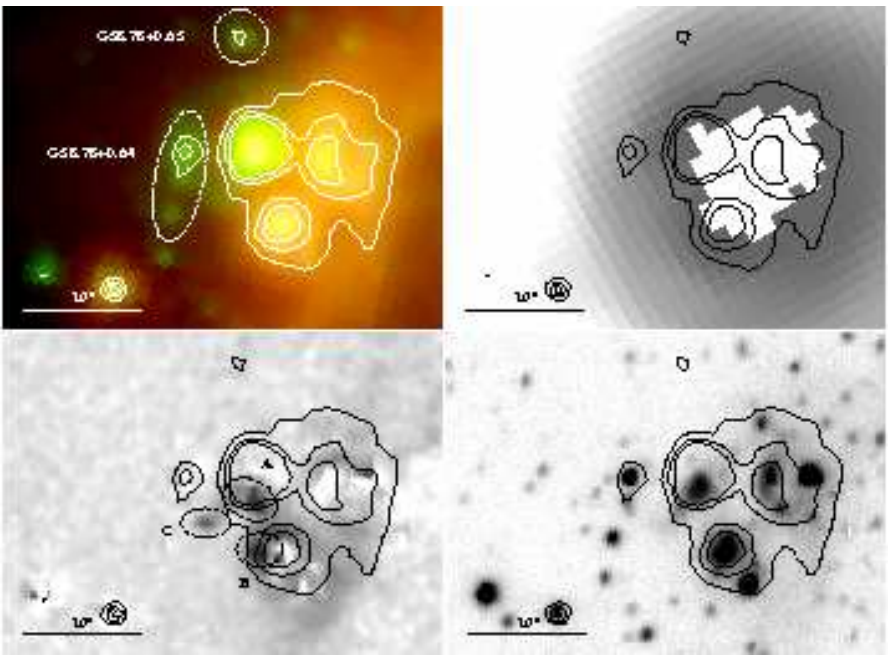}
\caption{Same as Figure~\ref{fig:G11.11-0.11}, but for EGO G58.78+0.64 and G58.78+0.65.  The positions of the two EGOs are marked by white dashed ellipses.}
\label{fig:G58.78+0.64}
\end{figure}

\begin{figure}
\includegraphics[angle=0,scale=0.65]{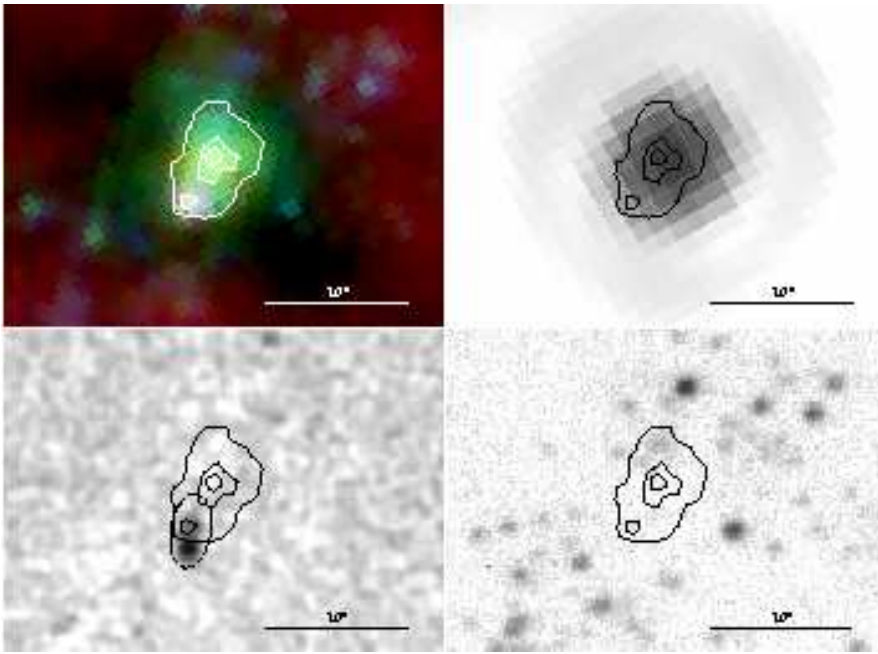}
\caption{Same as Figure~\ref{fig:G11.11-0.11}, but for EGO G59.79+0.63.  The black ellipse marks the H$_{2}$ lobe (MHO 2625).}
\label{fig:G59.79+0.63}
\end{figure}

\begin{figure}
\includegraphics[angle=0,scale=0.65]{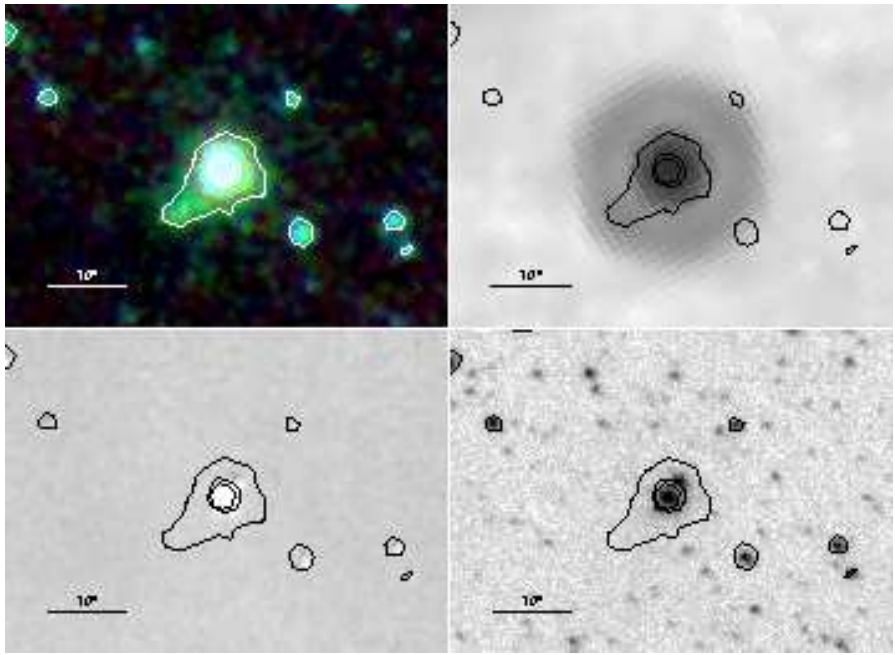}
\caption{Same as Figure~\ref{fig:G11.11-0.11}, but for EGO G62.70-0.51.}
\label{fig:G62.70-0.51}
\end{figure}

Of our 60 EGO sample, 39 exhibit H$_{2}$ and/or $K$-band detections, as are shown in Figures~\ref{fig:G11.11-0.11}-\ref{fig:G62.70-0.51}.  These figures include images in the IRAC bands (3.6, 4.5, and 8.0~$\micron$), MIPS 24~$\micron$, continuum-subtracted H$_{2}$, and $K$-band, representing the morphologies of the EGOs, the position of the protostars, the morphologies of the H$_{2}$ outflows, and $K$-band continuum emission, respectively.  The details of the 39 EGOs are described individually in the Appendix.  

Table~\ref{tab:h2} summarizes the detections of the H$_{2}$, $K$-, and $H$-band emission for the 60 EGOs.  Of the 60 EGOs, those labeled as ``Y'' and ``N'' are the EGOs with and without detections, respectively, in H$_{2}$, $K$-, and $H$-band emission.  Those EGOs with outflow candidates are labeled as ``Y?''.  In addition, we searched for the distances of the 60 EGOs in the literature, and listed the distances and the references in the Table.  The distances of the EGOs come from \citet{che09}, \citet{cyg09}, and \citet{he12}, using the same galactic rotation curve to estimate their kinematic distances \citep{rei09}.  


\subsection{H$_{2}$ Emission}

In this paper, we found 17 EGOs associated with H$_{2}$ outflows in the continuum-subtracted H$_{2}$ images of the 60 EGOs.  The H$_{2}$ outflow detection rate (28\%, 17/60) is slightly lower than that of Paper I (35\%, 12/34).  We will discuss this issue in Section 4.2.

Of the 17 EGOs with H$_{2}$ outflows, 10 EGOs are associated with an extended bipolar H$_{2}$ structure (EGO G12.91-0.26, G14.33-0.64, G21.24+0.19, G22.04+0.22, G24.33+0.14, G27.97-0.47, G29.96-0.79, G39.10+0.49, G48.66-0.30, and G49.42+0.33), and 7 EGOs exhibit H$_{2}$ lobes (EGO G11.11-0.11, G23.82+0.38, G34.39+0.22, G35.03+0.35, G37.48-0.10, G40.28-0.22, and G59.79+0.63).  All of the sources mentioned above are classified as sources with H$_{2}$ outflows.  In general, the H$_{2}$ outflows associated with these objects are highly collimated.  Molecular hydrogen emission-line object \citep[MHO,][]{davis10}\footnote{http://www.astro.ljmu.ac.uk/MHCat/} numbers are assigned for the newly identified H$_{2}$ outflows in Table~\ref{tab:mho}.

There are 12 EGOs which only show an isolated H$_{2}$ knot or extended H$_{2}$ emission (EGO G17.96+0.08, G19.36-0.03, G24.11-0.18, G24.33+0.14, G24.63+0.15, G34.26+0.15, G40.28-0.27, G53.92-0.07, G54.11-0.08, G57.61+0.02, G58.78+0.64, and G58.78+0.65).  The driving sources of the isolated H$_{2}$ knots are unclear, so they are classified as sources with candidate H$_{2}$ outflows.


For the 29 EGOs with H$_{2}$ emission detections, we only found five EGOs showing similar morphologies between the H$_{2}$ and 4.5~$\micron$ emission, (EGO G27.97-0.47, G34.39+0.22, G37.48-0.10, G49.42+0.33, and G53.92-0.07).  For the remaining 24 EGOs, the morphologies of the H$_{2}$ and 4.5~$\micron$ emission are different.  Generally, the H$_{2}$ emission are more extended than the 4.5~$\micron$ emission.


In the continuum-subtracted H$_{2}$ images, some of the EGOs show negative valued features, (e.g., EGO G12.91-0.26, G17.96+0.08, G24.11-0.17, G40.28-0.22).  These features probably represent the continuum emission with a large infrared excess and high extinction (Paper I).  The morphologies of the negative valued features usually resemble those of the $K$-band continuum emission.

\subsection{Near-infrared Continuum Emission}

In Table~\ref{tab:h2}, there are 32 EGOs showing extended $K$-band emission.  We compared the morphologies of the $K$-band and 4.5~$\micron$ emission for these EGOs, and found that at least six of them show similar morphologies between the two bands (EGO G12.91-0.26, G21.24+0.19, G40.60-0.72, G45.47+0.05, G50.36-0.42, and G57.61+0.02).  

\begin{figure}
\includegraphics[angle=0,scale=0.8]{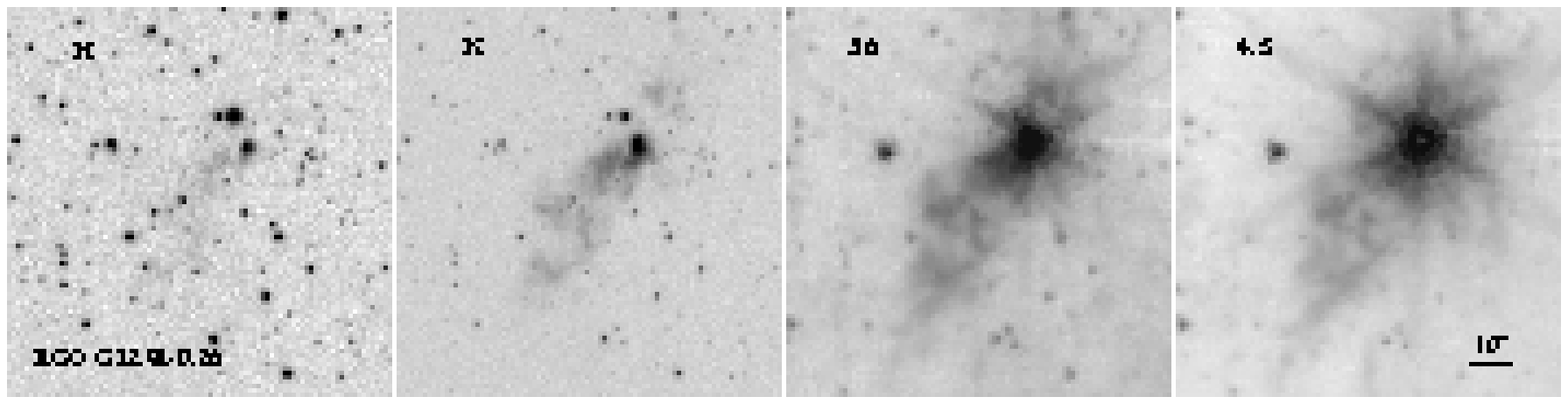}
\includegraphics[angle=0,scale=0.8]{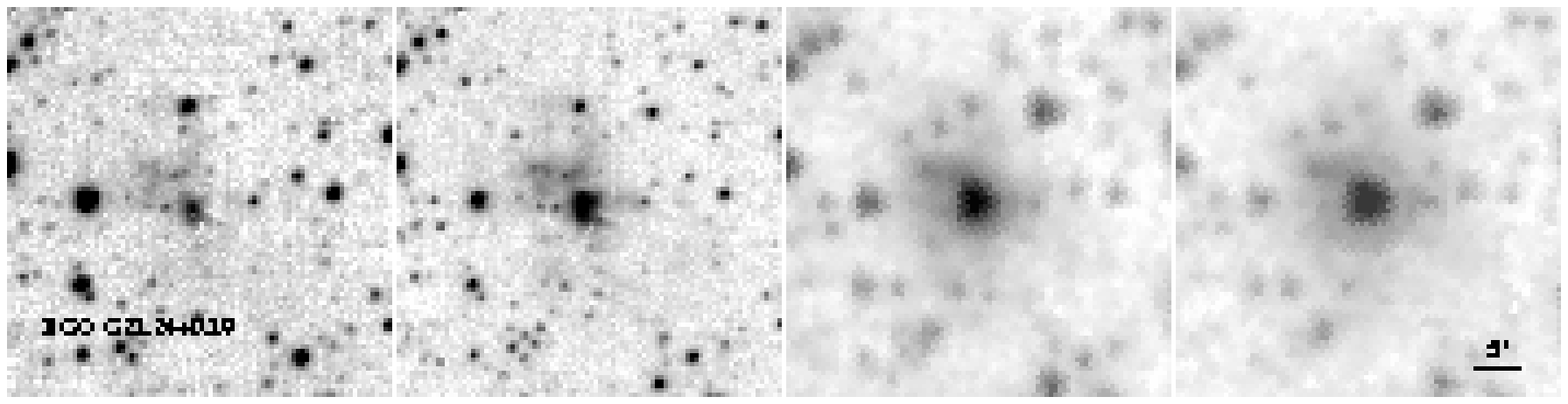}
\includegraphics[angle=0,scale=0.8]{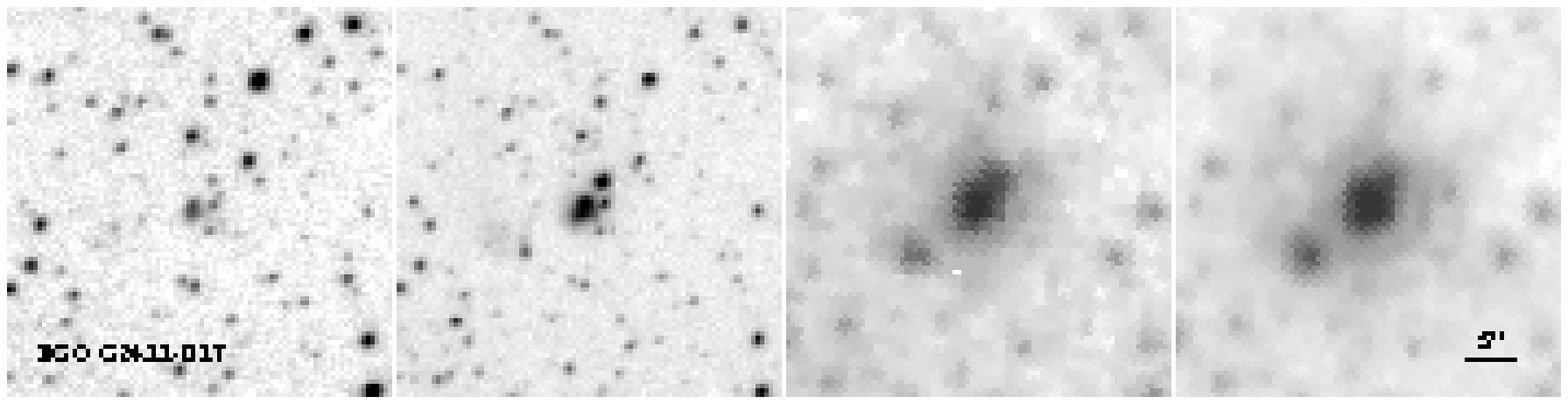}
\includegraphics[angle=0,scale=0.8]{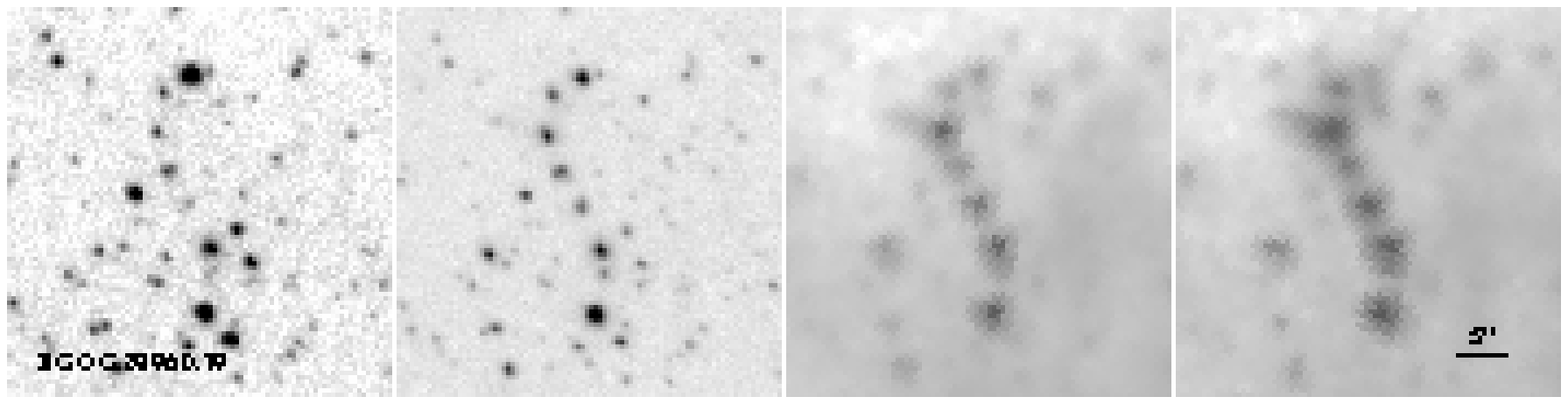}
\includegraphics[angle=0,scale=0.8]{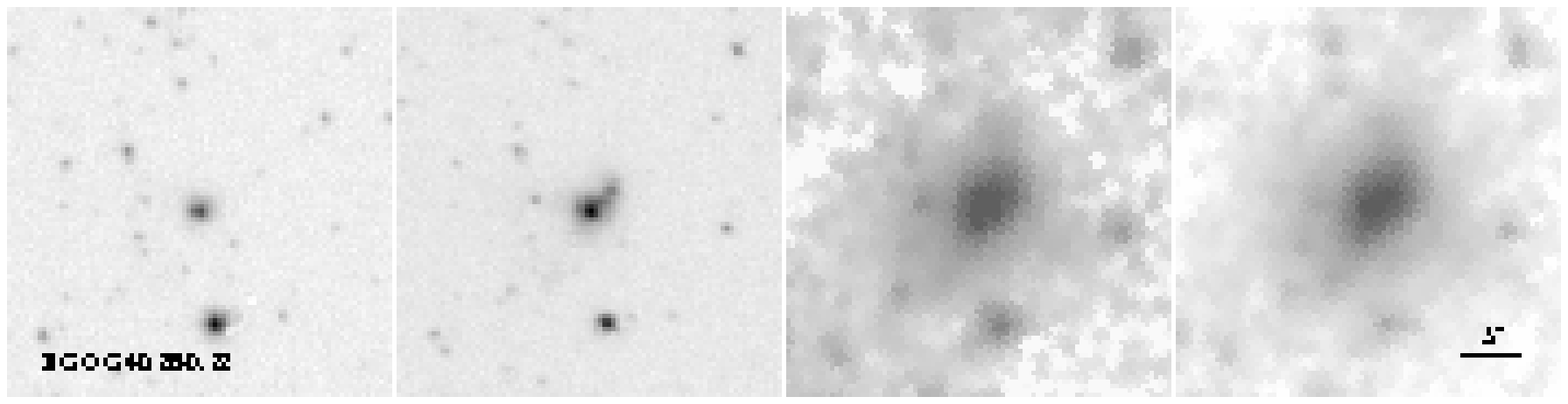}
\caption{EGOs with $H$-band emission; $H$, $K$, 3.6~$\micron$, and 4.5~$\micron$ images in the first, second, third, and fourth columns, respectively.}
\label{fig:continuum}
\end{figure}

There are 10 EGOs exhibiting $H$-band emission (Table~\ref{tab:h2}).  Figure~\ref{fig:continuum} shows the morphologies of the 10 EGOs in the $H$, $K$, 3.6~$\micron$, and 4.5~$\micron$ bands. As seen in the 7 objects in Paper~I, the flux distribution changes gradually with wavelength.  In addition, the morphologies in different bands are similar for 5 objects (EGO G24.11-0.17, G29.96-0.79, G45.47+0.13, G53.92-0.07, and G62.70-0.51), except for the effect of different angular resolutions.

Compared with the H$_{2}$ emission of the EGOs, the morphologies of the $K$-band diffuse emission are more similar to that seen at the 4.5~$\micron$.  Furthermore, for those EGOs with $H$-band detections, the morphologies of the $H$ and $K$-band emission resemble each other.  These results are consistent with Paper~I.

\subsection{Methanol Master Distribution}

Class I methanol masers have been suggested as an outflow tracer, and observations have found a correlation between methanol masers and H$_{2}$ outflows, e.g., IRAS 16547-4247 \citep{vor06}, IRAS 20126+4104 \citep{kur04}, Orion KL \citep{joh92}, and DR 21 \citep{pla90}.  \citet{cyg09} performed a Class I 44~GHz methanol maser survey for 19 EGOs in the northern Galactic plane using the VLA, and 17 of them show positive detections.  This survey provides an additional sample for follow-up studies.  

All of the 19 EGOs with Class I 44~GHz methanol maser observations \citep{cyg09} are covered by UWISH2, so we compare the 44~GHz observation results with our H$_{2}$ observations.  Combined with the Paper I results, we provide the largest data set of H$_{2}$ observations for Class I 44~GHz methanol masers.  We found that there are 8 EGOs with both H$_{2}$ outflow detections and 44~GHz methanol maser observations (i.e., EGO G11.92-0.61, G19.01-0.03, G19.36-0.03, G22.04+0.22, G35.03+0.35, G37.48-0.10, G39.10+0.49, and G49.42+0.33), as described in the Appendix.  Figures~\ref{fig:11.92-0.61}-\ref{fig:39.10+0.49} show the continuum-subtracted H$_{2}$ images and the distributions of those EGOs with 44~GHz methanol masers detections.

We found that the 44~GHz methanol masers can be located upstream or/and downstream of the H$_{2}$ lobes in the continuum-subtracted H$_{2}$ images.  The H$_{2}$ lobes could be associated with primarily blueshifted 44~GHz methanol masers (e.g., EGO G11.92-0.61, G19.01-0.03, and G35.03+0.35), or with redshifted only (EGO G39.10+0.49).  Some of the H$_{2}$ lobes show both blueshifted and redshifted masers (e.g., EGO G19.36-0.03, G22.04+0.22, and G37.48-0.10).

We also found that some bright H$_{2}$ spots are not associated with 44~GHz methanol masers, e.g., EGO G11.92-0.61 and G39.10+0.49.  For the case of EGO G49.42+0.33, there is no 44~GHz methanol maser detection \citep{cyg09}.  However, we found that the EGO was associated with an H$_{2}$ outflow (Figure~\ref{fig:G49.42+0.33}).


\subsection{Radio Detections from the Literature}

We searched the literature for radio detections for the 94 EGOs, and found 18 EGOs associated with radio sources\footnote{EGO G35.03+0.35 harbors at least two radio sources.  CM1 is a UC HII region and CM2 is a HC HII region candidate \citep{cyg11}.}.  Seven EGOs are associated with UC~\ion{H}{2} regions, 10 EGOs are associated with HC \ion{H}{2} regions or candidates, and two EGOs are associated with radio thermal jets.  Table~\ref{tab:radio} lists these EGOs and the references.

\section{DISCUSSION}

In this section, we discuss the 94 EGOs, including the 34 EGOs from Paper~I, and their H$_{2}$ outflows, continuum emission, and radio detections.  Besides discussing the general properties of the EGOs, we also consider two subcatalogs: ``likely'' and ``possible'' MYSO outflow candidates, as divided by \citet{cyg08}.  The differences between ``likely'' and ``possible'' sources are that the 4.5~$\micron$ emission of the ``possible'' sources could be confused by multiple nearby point sources or images artifacts from near bright sources.

\subsection{H$_{2}$ Outflows}

\begin{figure}
\includegraphics[angle=90,scale=1.1]{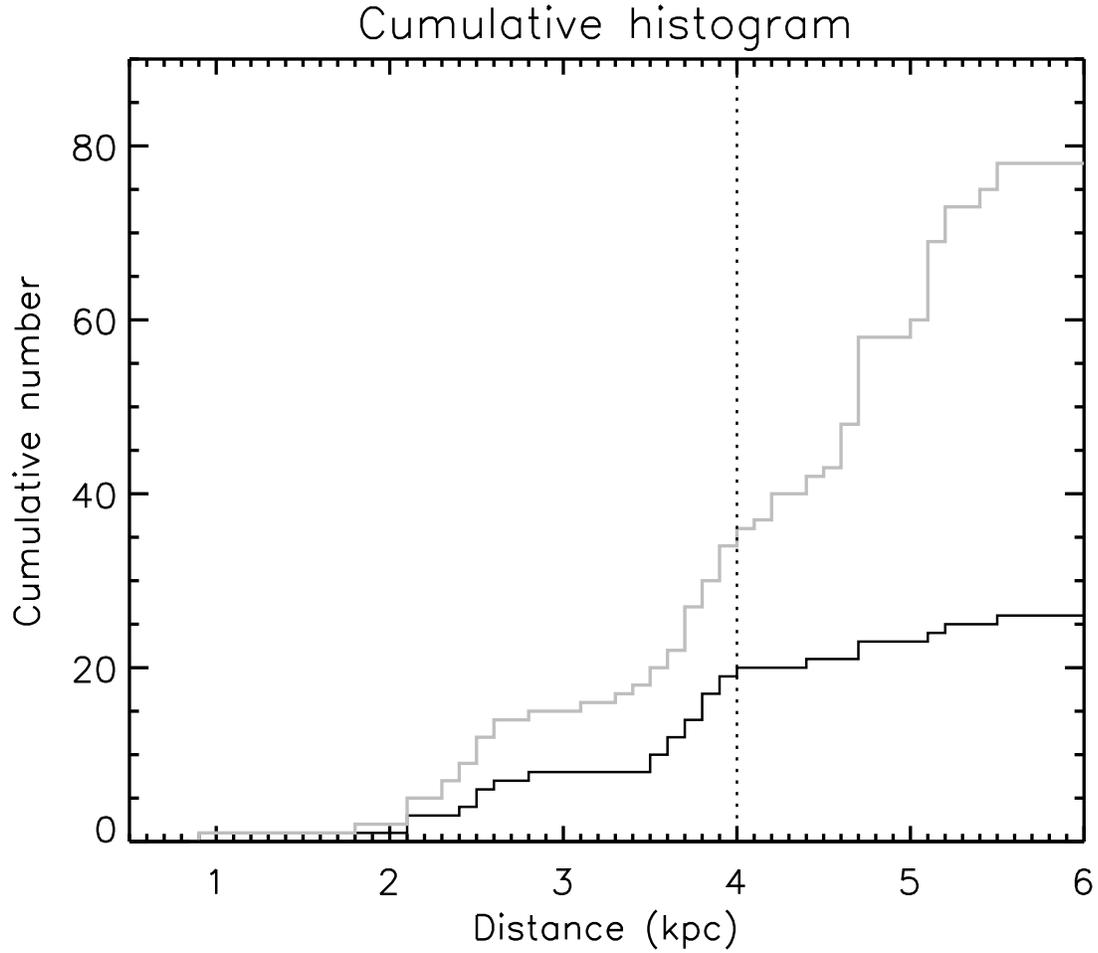}
\caption{Cumulative histograms for EGOs with H$_{2}$ outflows (black) and the entire EGO sample (gray) with bin size 0.1~kpc.  The cumulative histograms are similar for those EGOs within 4~kpc, and the ratio of the cumulative numbers of those EGOs with H$_{2}$ outflows to the entire EGO sample is $\sim$0.5.  For those EGO with distances beyond 4~kpc, the cumulative number of EGOs with outflows becomes flat.  However, the cumulative number of the entire EGOs sample still increases.}
\label{fig:distance}
\end{figure}

There are 82 EGOs with known kinematic distances (28 with H$_{2}$ outflows; 13 with outflow candidates; 41 without outflow detections).  Figure~\ref{fig:distance} shows the cumulative distance histograms within 6~kpc for those EGOs with H$_{2}$ outflows and the entire EGO sample.  For those EGOs within 4~kpc, the distributions of the two cumulative histograms are almost identical, based on the Kolmogorov-ÐSmirnov test \citep[probability is 0.99;][]{pre07}.  The probability drops to 0.04, if considering all of the EGOs within 6~kpc.  In the Figure, there are only a few EGOs within $\sim$2~kpc, and the numbers increase for both cumulative histograms beyond $\sim$2~kpc.  Between $\sim$2.8 and $\sim$3.5~kpc, the cumulative numbers become flat, and then increase again after $\sim$3.5~kpc.  The distributions may represent the presence of spiral arms, where the cumulative numbers increase from $\sim$2 to $\sim$2.8~kpc and beyond $\sim$3.5~kpc.

The similarity between the two cumulative histograms also suggests that the detection rates are similar within 4~kpc.  The ratio of those EGOs with H$_{2}$ outflows to the entire EGOs sample is $\sim$50\%, and the detection rate becomes lower beyond 4~kpc.  Thus, it is reasonable to claim that about half EGOs are associated with H$_{2}$ outflows, based on the sensitivity of UWISH2.



\citet{ioa12b} used UWISH2 images to search for H$_{2}$ outflows along the Galactic plane in Serpens and Aquila, and identified 131 outflows from YSOs.  The distances of these H$_{2}$ outflows are estimated from foreground star counts, and are in the range of 2 to 5 kpc.  They suggested that a detection limit of UWISH2 observations of $\sim$5~kpc.  Our results are generally consistent with theirs, except for EGO G49.42+0.33.  The kinematic distance of EGO G49.42+0.33 is around 12.3~kpc \citep{che09,cyg09,he12}.  However, this EGO is detected to be associated with a bipolar H$_{2}$ outflow.  Based on the detection limit estimation by \citet{ioa12b} and our results, we are probably unable to detect H$_{2}$ outflow at that distance.  The molecular cloud, where the EGO is located, may have peculiar velocity, resulting in an overestimate of its kinematic distance.  Another possibility is that the H$_{2}$ outflow is more extended and brighter than the typical H$_{2}$ outflows, so it can be detected at the distance.



The excitation conditions of the H$_{2}$ line emission in the 4.5~$\micron$-band is similar to $\upsilon$=1-0 $S$(1) at 2.122~$\micron$ \citep{tak10}.  One would expect to see similar morphologies between the 4.5~$\micron$ and the H$_{2}$ emission, if the 4.5~$\micron$ emission really comes from the H$_{2}$ line emission.  However, the morphologies between H$_{2}$ and 4.5~$\micron$ emission are different for most of the EGOs with H$_{2}$ outflows.

On the other hand, 8 EGOs ($\sim$28\% of EGOs with H$_{2}$ emission) show similar morphologies between H$_{2}$ and 4.5~$\micron$ emission.  This also can be seen in the energetic outflow from DR 21 \citep{mar04}.  The similar morphologies between H$_{2}$ and 4.5~$micron$, implying that 4.5~$\micron$ emission likely comes from H$_{2}$ emission.  However, these sources comprise less than one third of the EGOs with H$_{2}$ outflows.

In addition, some previous observations support the premise that the 4.5~$\micron$ emission comes from H$_{2}$ emission, e.g., 6 outflows from low-mass YSOs from nearby star-forming regions \citep{tak10}.  We used these 6 outflows to investigate their properties.  As described in \citet{cyg08}, the faintest EGOs from GLIMPSE have surface brightness $\gtrsim$ 4~MJy~sr$^{-1}$.  The 4.5~$\micron$ surface brightnesses of the 6 outflows are generally less than 4~MJy~sr$^{-1}$, except for some bright knots.  The sizes of the bright knots are $\lesssim 10\arcsec$.  If they were located a few kpc away, then they would not be spatially resolved by $Spitzer$.  In summary, the extended features of the 6 outflows from low-mass YSOs are undetectable in GLIMPSE.




The H$_{2}$ outflow detection rates for the ``likely'' MYSO outflow candidates (42\%) are higher than those for the ``possible'' ones (19\%) for the entire sample.  If only considering those EGOs within 4~kpc, the rates increase to 62\% for the former and 44\% for the later.  This suggests that some of the ``possible'' candidates could be really affected by multiple nearby point sources or images artifacts from near bright sources, and they may be not bright at 4.5~$\micron$ intrinsically.  Another possibility is the ``possible'' MYSO outflow candidates associated with weaker outflow activities, implying these two subcategories may be in different evolution stages.


\subsection{Morphologies of H$_{2}$ and Infrared Continuum Emission}

The majority of the $K$ and $H$-band diffuse continuum emission comes from the scattered continuum from embedded YSOs in star-forming regions \citep{con07,hod94}.  The surrounding gas or the outflow cavities can scatter the emission from YSOs \citep{has08,has07,tam06}.  This can also be seen in EGOs, e.g., the polarization observation of EGO G35.20-0.74 \citep{wal90} and the adaptive-optics-assisted near-infrared integral field spectroscopic observations of EGO G12.91-0.26 (W33A) \citep{dav10}.


The entire $K$-band detection rate of the EGOs is 57\% (54/94).  For the ``likely'' and ``possible'' MYSO outflow candidates, their $K$-band detection rates are similar, 56\% and 60\%, respectively. 
Based on the UKIDSS GPS observations, the $H$-band detection rate for the EGOs is 19\% (18/94) in the northern Galactic plane.  

The EGOs with $K$-band emission have a significantly higher detection rate for H$_{2}$ outflows.  For those EGOs with $K$-band detections and distances $\leq4$~kpc, the H$_{2}$ outflow detection rate is 61\%.  In contrast, for those EGO without $K$-band detections within 4~kpc, the H$_{2}$ outflow detection rate is 36\%.  Since the $K$-band emission comes from the scattered continuum from the outflow cavity or the surrounding gas of an embedded YSO, it is natural to have a correlation between $K$-band and H$_{2}$ outflow detections.  The correlation can explain why the H$_{2}$ outflow detection rate in Paper I is slightly higher than that in this work, as described in Section 3.1.  The differences are due to the sample in Paper I including some preselected EGOs with $K$-band detections.

Compared with the H$_{2}$ emission, the morphologies of the $K$-band continuum emission are usually more compact and closer to the driving sources \citep{lee12,dav07,var10}.  The similar morphologies between the $K$-band and 4.5~$\micron$ emission imply the same origin for these two infrared emission.  We suggest that the 4.5~$\micron$ emission primarily comes from scattered continuum from the embedded YSOs, with a smaller contribution from the H$_{2}$ line emission.  The 4.5~$\micron$ emission is unlikely to represent the H$_{2}$ line emission in the 4.5~$\micron$ band, as the $K$-band continuum emission can not represent the H$_{2}$ line emission at 2.122~$\micron$.

\subsection{Class I 44 GHz Methanol Maser and Radio Continuum Detections}


Class I methanol masers are collisionally excited, and were suggested as an outflow tracer.  As described in Section 3.3, there is a variety of distributions between the H$_{2}$ outflows and the Class I 44~GHz methanol masers, which can be located upstream or downstream of the H$_{2}$ emission.  Some bright H$_{2}$ spots or outflows are even not associated with any methanol maser.  It could be that the methanol masers and H$_{2}$ emission trace different excitation conditions.


In order to determine if UC and HC \ion{H}{2} regions are still accreting or not, we use the existences of H$_{2}$ outflows as an indicator.  Although this is not a direct measurement of mass accretion, there is growing evidence that outflow activities are powered by mass accretion \citep{dav04,cal97,har95}.  The former is much easier to observe.

Four of the 10 HC \ion{H}{2} regions are associated with H$_{2}$ outflows (EGO G11.92-0.61, G12.91-0.26, G35.03+0.35, and G35.20-0.74).  If we consider those EGOs within 4~kpc, then the H$_{2}$ detection rate is 57\% (4/7), which is similar to that of the EGOs.  It implies that the majority of the MYSOs in the HC~\ion{H}{2} regions are still accreting.

For the 7 EGOs associated with UC \ion{H}{2} regions, only one EGO (G24.33+0.14) is associated with an H$_{2}$ outflow (Figure~\ref{fig:G24.33+0.14}).  However, of the 7 UC \ion{H}{2} regions, only two are located within 4~kpc, so the sample is too limited to conclude if UC \ion{H}{2} regions are accreting or not.


For UC and HC \ion{H}{2} regions, their radio sources coincide with the positions of the massive stars.  However, the two radio sources (IRAS 18182-1433b and 18264-1152b) do not coincide with the 24~$\micron$ sources of EGO G16.59-0.05 and G19.88-0.53, respectively, and their radio spectral indexes are consistent with thermal jets.  Furthermore, these two radio sources are associated with H$_{2}$ outflows.  Throughout, the radio emission of IRAS 18182-1433b and 18264-1152b is contributed by thermal jets.



\citet{cyg11} performed deep radio observations of EGOs with the VLA, and found that 8 out of 14 EGOs did not show any radio emission.  Of the 8 radio quiet EGOs, three EGOs are associated with H$_{2}$ outflows and one with H$_{2}$ an outflow candidate.  Since these source are still accreting, the infall gas would absorb the stellar UV photons and quench the \ion{H}{2} regions \citep{wal95,ket07}.  This would reduce the radio detection rate, even though the YSOs are massive enough to produce \ion{H}{2} regions.  

\section{CONCLUSIONS}

In this paper, we present the narrowband H$_{2}$ (UWISH2) and near-infrared continuum (UKIDSS GPS) results from UKIRT for 60 EGOs.  In general, we find that the morphologies of the {\it Spitzer} 4.5~$\micron$ emission are generally similar to those of the near-infrared continuum, but different from those of the H$_{2}$ emission.  The 4.5~$\micron$ emission is seldom associated with 2.122~$\micron$ line H$_{2}$ emission ($\sim$28\% of those EGOs with H$_{2}$ emission).  We suggest that the 4.5~$\micron$ emission comes mainly from scattered continuum and partially from H$_{2}$ emission.  The 4.5~$\micron$ emission is unlikely to represent the H$_{2}$ line emission at 4.5~$\micron$ band, as the $K$-band continuum emission does not represent the H$_{2}$ line emission at 2.122~$\micron$.  These results are consistent with our previous studies.

Combined with our previous studies, we present all 94 EGOs in the northern Galactic plane.  We summarize our findings below.  

\begin{enumerate}

\item We found that our detection of H$_{2}$ emission for EGOs is complete to a distance of about 4~kpc based on the sensitivity of UWISH2 observations, and about half of EGOs are associated with H$_{2}$ outflows.


\item When considering those EGOs within 4~kpc, the H$_{2}$ outflow detection rate for those EGOs classified as ``likely'' MYSO outflow candidates (62\%) is higher than that for ``possible'' MYSO outflow candidates (44\%).  The lower detection rate for the later may be due to the 4.5~$\micron$ excesses being artifacts from nearby multiple point or bright sources, or they are in different evolution stages.

\item There is a variety of distributions between the Class I 44 GHz methanol masers with respect to the H$_{2}$ outflows. The masers could be located upstream or downstream of the H$_{2}$ emission.  Some bright H$_{2}$ spots or outflows are not associated with methanol masers.  It could be that the methanol masers and H$_{2}$ emission trace different excitation conditions.

\item There is a positive correlation between the $K$-band and H$_{2}$ outflow detections.  For those EGOs within 4~kpc, the H$_{2}$ outflow detection rate for the EGOs with $K$-band detections (61\%) is significantly higher than that of the EGOs without $K$-band detections (36\%), respectively.  Since the $K$-band emission comes from the scattered continuum of the the outflow cavity or the surrounding gas of an embedded YSO, it is natural to see the correlation.

\item Based on a literature search, there are 7 EGOs associated with HC \ion{H}{2} regions.  More than half of the HC \ion{H}{2} regions have companion H$_{2}$ outflows, suggesting that at least half of them are still accreting.
\end{enumerate}



\acknowledgments

We thank the anonymous referee for constructive comments and suggestions.  This research made use of the SIMBAD data base operated at CDS, Strasbourg, France, and NASA's Astrophysics Data System Abstract Service.  MT is supported from National Science Council of Taiwan (Grant No. NSC-100-2112-M-001-007-MY3).

\appendix

\section{Selected Individual Sources}\label{appendix}

We present those EGOs with H$_{2}$ and/or $K$-band emission detections.

{\it EGO G11.11-0.11 (Figure~\ref{fig:G11.11-0.11})}.  The 4.5~$\micron$ emission surrounds a 24~$\micron$ source, and the lowest contour around the EGO roughly represents its shape.  Two aligned H$_{2}$ knots are identified in the continuum-subtracted H$_{2}$ image; the 24~$\micron$ source could be its driving source.  A $K$-band diffuse source is found around the 4.5 contour peak.

{\it EGO G12.91-0.26 (Figure~\ref{fig:G12.91-0.26})}.  EGO G12.91-0.26 is associated with a radio source (W33A) with spectral index $\sim1.2\pm0.4$ \citep{van05}, suggesting that W33A could be a HC \ion{H}{2} region.  A bright saturated 24~$\micron$ source is located at the 4.5~$\micron$ contour peak.  The main 4.5~$\micron$ emission extends toward the SE direction.  Based on adaptive-optics-assisted near-infrared integral field spectroscopic observations, \citet{dav10} suggested that the $K$-band emission is light reflected from the YSO W33A.  They found an H$_{2}$ outflow very close to the driving source and the outflow extending along the SE direction (the same as the 4.5~$\micron$ emission).  In our continuum-subtracted H$_{2}$ image, we did not find the H$_{2}$ outflow identified by \citet{dav10}, because the H$_{2}$ outflow is located very close to the MYSOs where the scattered continuum dominates the near-infrared emission.  However, we found another bipolar H$_{2}$ outflow in the E-W direction.  Its orientation is different from that of the 4.5~$\micron$ emission.  In the $K$-band image, its morphology is similar to that of the 4.5~$\micron$ emission.

{\it EGO G14.33-0.64 (Figure~\ref{fig:G14.33-0.64})}.  The 4.5~$\micron$ emission extends beyond the \ion{H}{2} region \citep[G14.33-0.64,][]{sew04}, which shows a reddish color in the $Spitzer$ image.  There are two 24~$\micron$ sources.  One is related to the \ion{H}{2} region, and the other one may be associated with the EGO.  Two H$_{2}$ lobes (A and B) and one H$_{2}$ knot (C) can be seen in the continuum-subtracted H$_{2}$ image; they do not coincide with the EGO.  The two H$_{2}$ lobes and the 24~$\micron$ source associated with the EGO are aligned, so the 24~$\micron$ source could be the driving source of the two H$_{2}$ lobes.

{\it EGO G17.96+0.08 (Figure~\ref{fig:G17.96+0.08})}.  The 4.5~$\micron$ emission is distributed around a 24~$\micron$ source.  An H$_{2}$ knot is located at a distance of $\sim23\arcsec$.  It is not clear if the H$_{2}$ knot is physically related to the EGO or not.  There is a $K$-band source at the peak of the 4.5~$\micron$ contours.

{\it EGO G19.36-0.03 (Figure~\ref{fig:G19.36-0.03})}.  The 4.5~$\micron$ emission extends toward the SE.  There is a 24~$\micron$ source located at the peak of the 4.5~$\micron$ emission.  Faint H$_{2}$ emission is found in the continuum-subtracted H$_{2}$ image, and extends along the NE-SW direction.  It is unclear if the 24~$\micron$ source is the driving source or not.  An extended $K$-band source is located at the peak of the 4.5~$\micron$ contour map.

{\it EGO G21.24+0.19 (Figure~\ref{fig:G21.24+0.19})}.  A 24~$\micron$ source is located at the peak of the 4.5~$\micron$ emission.  There is a bipolar H$_{2}$ outflow identified by \citet{ioa12a} (MHO 2255 and 2256).  The 24~$\micron$ source could be the driving source.  There is a $K$-band extended source coincident with the EGO.

{\it EGO G22.04+0.22 (Figure~\ref{fig:G22.04+0.22})}.  There is a 24~$\micron$ source at the 4.5~$\micron$ contour peak.  A bipolar H$_{2}$ outflow (MHO 2260) has been reported by \citet{ioa12a} with the 24~$\micron$ source, which is likely the driving source of the outflow.  No $K$-band source is found in the $K$-band image.

{\it EGO G23.82+0.38 (Figure~\ref{fig:G23.82+0.38})}.  There is a 24~$\micron$ source located at the 4.5~$\micron$ contour peak with the EGO extending to the NW of the peak.  \citet{ioa12a} identified an H$_{2}$ lobe (MHO 2262).  The 24~$\micron$ source could be the driving source of the outflow.  In the $K$-band image, there is a point source at the 4.5~$\micron$ peak.

{\it EGO G24.11-0.17 and G24.11-0.18 (Figure~\ref{fig:G24.11-0.17})}.  There are two EGOs in the figure.  Each is associated with a 24~$\micron$ source.  There is an H$_{2}$ knot near and slightly extended toward G24.11-0.18, so the H$_{2}$ knot is more likely related to the EGO.  In the $K$-band image, there are multiple sources around the G24.11-0.17, and one source associated with G24.11-0.18.  


{\it EGO G24.33+0.14 (Figure~\ref{fig:G24.33+0.14})}.  An \ion{H}{2} region \citep[G24.33+0.11 GLM1,][]{bat10} is found at the peak of the 4.5~$\micron$ emission.  There is a 24~$\micron$ source located at the 4.5~$\micron$ contour peak.  In the continuum-subtracted H$_{2}$ image, an H$_{2}$ outflow (MHO 3258) with two H$_{2}$ knots are identified.  There is no $K$-band source at the 4.5~$\micron$ contour peak.

{\it EGO G24.63+0.15 (Figure~\ref{fig:G24.63+0.15})}.  At the 4.5~$\micron$ contour peak, there is a 24~$\micron$ source.  An H$_{2}$ knot is found in the continuum-subtracted H$_{2}$ image, but it is unclear if the H$_{2}$ knot is related to the EGO or not.  In the $K$ image, a faint $K$-band source is located at the 4.5~$\micron$ contour peak.

{\it EGO G27.97-0.47 (Figure~\ref{fig:G27.97-0.47})}.  A 24~$\micron$ source is located near the peak of the 4.5~$\micron$ contour.  There is an H$_{2}$ outflow (MHO 2441) reported by \citep{ioa12a}.  Its morphology is similar to that of the 4.5~$\micron$ emission.  In $K$-band image, the extended emission coincides with the western H$_{2}$ lobe.

{\it EGO G29.96-0.79 (Figure~\ref{fig:G29.96-0.79})}.  The EGO is elongated along the NS direction, and associated with several point sources.  There are two possible faint 24~$\micron$ sources, located in the north part of the EGO.  In the continuum-subtracted H$_{2}$ image, an H$_{2}$ outflow is identified and distributed along the EW direction.  It is unclear which point source is the driving source.  Each 4.5~$\micron$ peak has a corresponding $K$-band source in the $K$-band image.

{\it EGO G34.26+0.15 (Figure~\ref{fig:G34.26+0.15})}.  In this field, there is a cometary UC~\ion{H}{2} region G34.3+0.2 \citep{rei85} with its tail pointing toward the west.  In addition, there are two HC~\ion{H}{2} regions \citep[G34.26+0.15~A and B,][]{gau94,sew11} located near the head of the cometary UC~\ion{H}{2} region.  In the 24~$\micron$ image, the locations of the two HC~\ion{H}{2} regions are saturated.  Several H$_{2}$ knots are found in the continuum-subtracted H$_{2}$ image, but it is not clear if they are associated with the HC~\ion{H}{2} regions or not.  No $K$-band sources are found near the EGO.

{\it EGO G34.28+0.18 (Figure~\ref{fig:G34.28+0.18})}.  In the 4.5~$\micron$ contour map, there are two peaks, each of which is associated with a 24~$\micron$ source.  No H$_{2}$ emission is detected around the EGO.  In the $K$-band image, two fan shaped diffuse sources are found around the north 24~$\micron$ source.

{\it EGO G34.39+0.22 (Figure~\ref{fig:G34.39+0.22})}.  The EGO is elongated in the NS direction.  There is a 24~$\micron$ source located near the 4.5~$\micron$ contour peak.  In continuum-subtracted H$_{2}$ image, an H$_{2}$ lobe is identified, and the 24~$\micron$ source could be its driving source.  In general, the morphologies of the 4.5~$\micron$ and H$_{2}$ emission are similar.  No apparent $K$-band source is seen in the $K$-band continuum image.

{\it EGO G35.03+0.35 (Figure~\ref{fig:G35.03+0.35})}.  \citet{cyg11} found several radio centimeter sources associated with the EGO.  EGO G35.03+0.35~CM1 exhibits free-free emission.  EGO G35.03+0.35~CM2 is associated with 6.7~GHz methanol masers \citep{cyg09} and could be a HC~\ion{H}{2} region \citep{cyg11}.  The 4.5~$\micron$ emission of EGO~G35.03+0.35 is extended in the NE-SW direction.  A saturated 24~$\micron$ source is located near the 4.5~$\micron$ contour peak.  In the continuum-subtracted H$_{2}$ image, an H$_{2}$ lobe is found, and likely extends from G35.03+0.35~CM2.  The $K$-band emission is elongated along the NE-SW direction, similar to the 4.5~$\micron$ emission.

{\it EGO G37.48-0.10 (Figure~\ref{fig:G37.48-0.10})}.  The EGO is elongated along the EW direction.  There is a 24~$\micron$ source located near the peak of the 4.5~$\micron$ contours.  In the continuum-subtracted H$_{2}$ image, an H$_{2}$ outflow lobe is distributed toward the west of the EGO.  A $K$-band diffuse source is found around the 4.5 contour peak.

{\it EGO G39.10+0.49 (Figure~\ref{fig:G39.10+0.49})}.  The EGO is distributed along the NW-SE direction.  There is a 24~$\micron$ source located at the 4.5~$\micron$ contour peak.  In the continuum-subtracted H$_{2}$ image, a bipolar H$_{2}$ outflow is identified with a morphology different from that of the 4.5~$\micron$ emission.  A $K$-band extended source is located at the peak of the 4.5~$\micron$ contour map, and the $K$-band emission also extends along the NW-SE axis.

{\it EGO G39.39-0.14 (Figure~\ref{fig:G39.39-0.14})}.  The 4.5~$\micron$ contour peak coincides with a 24~$\micron$ source.  In addition, an \ion{H}{2} region \citep[IRAS 19012+0536 B,][]{hof11} is also located near the peak.  There is no H$_{2}$ emission found in the continuum-subtracted H$_{2}$ image.  In the $K$-band image, a $K$-band diffuse source extends along the axis of the EGO.

{\it EGO G40.28-0.22 (Figure~\ref{fig:G40.28-0.22})}.  A saturated 24~$\micron$ source is surrounded by the 4.5~$\micron$ emission.  There is an H$_{2}$ lobe extending from the 24~$\micron$ source in the continuum-subtracted H$_{2}$ image.  An extended $K$-band source is located at the 4.5~$\micron$ contour peak.

{\it EGO G40.28-0.27 (Figure~\ref{fig:G40.28-0.27})}.  A 24~$\micron$ source is located at the 4.5~$\micron$ contour peak.  In the continuum-subtracted H$_{2}$ image, an H$_{2}$ knot is found.  It is unclear, if the H$_{2}$ knot is related to the EGO or not.  There is a faint $K$-band diffuse source at the peak of the 4.5~$\micron$ emission.

{\it EGO G40.60-0.72 (Figure~\ref{fig:G40.60-0.72})}.  In the 4.5~$\micron$ contour map, there are two contour peaks, the NW one associated with the EGO.  A saturated 24~$\micron$ source is located between the two 4.5~$\micron$ contour peaks.  No H$_{2}$ emission is found.  The SE and NW 4.5~$\micron$ peaks are associated with a $K$-band point source and an extended source, respectively.  The morphologies of the 4.5~$\micron$ and $K$-band emission are similar.

{\it EGO G45.47+0.05 (Figure~\ref{fig:G45.47+0.05})}.  The EGO is associated with an \ion{H}{2} region \citep[G045.4657+00.0453,][]{urq09} which is located around the peak of the saturated 24~$\micron$ source.  No H$_{2}$ emission is found in the continuum-subtracted H$_{2}$ image.  The $K$-band emission peaks at the the contour peak of the 4.5~$\micron$ emission contour.

{\it EGO G45.47+0.13 (Figure~\ref{fig:G45.47+0.13})}.  The EGO is located at the edge of an \ion{H}{2} region which saturates the 24~$\micron$ image.  No H$_{2}$ emission is found in the continuum-subtracted H$_{2}$ image.  There is $K$-band diffuse source at the peak of the 4.5~$\micron$ contours, and the morphologies of the $K$-band and 4.5~$\micron$ emission are similar.

{\it EGO G45.80-0.36 (Figure~\ref{fig:G45.80-0.36})}.  The EGO is associated with a 24~$\micron$ source.  In the continuum-subtracted H$_{2}$ image, there is no H$_{2}$ emission detection.  The $K$-band image shows a diffuse source around the 4.5~$\micron$ contour peak.

{\it EGO G48.66-0.30 (Figure~\ref{fig:G48.66-0.30})}.  EGO G48.66-0.30 is faint in the 4.5~$\micron$ image.  In the 4.5~$\micron$ contour map, there are two peaks.  The west peak is associated with a faint 24~$\micron$ source.  We found a bipolar H$_{2}$ outflow.  The morphologies of the H$_{2}$ and the 4.5~$\micron$ emission are different.  The 24~$\micron$ source could be the driving source of the H$_{2}$ outflow.  No $K$-band source is associated with the 4.5~$\micron$ emission.

{\it EGO G49.07-0.33 (Figure~\ref{fig:G49.07-0.33})}.  In the 4.5~$\micron$ contour peak, there is a 24~$\micron$ source.  No H$_{2}$ emission is associated with the EGO.  A diffuse $K$-band source coincides with the 4.5~$\micron$ peak.

{\it EGO G49.27-0.34 (Figure~\ref{fig:G49.27-0.34})}.  This EGO is associated with two 24~$\micron$ sources; each of them coincides with a radio source \citet{cyg11}.  In the continuum-subtracted H$_{2}$ image, there is no H$_{2}$ emission; however $K$-band continuum emission can be seen in the figure.  This supports the result of Germini $L$- and $M$-band spectral observations \citep{deb10}.  This EGO does not show H$_{2}$ emission and the infrared emission comes from the scattered continuum.

{\it EGO G49.42+0.33 (Figure~\ref{fig:G49.42+0.33})}.  The EGO is elongated along the N-S axis.  There is no apparent 24~$\micron$ source associated with this EGO.  An H$_{2}$ bipolar outflow is identified and is also elongated in the N-S direction.  One 4.5~$\micron$ peak coincides with $K$-band diffuse emission.

{\it EGO G50.36-0.42 (Figure~\ref{fig:G50.36-0.42})}.  In the 4.5~$\micron$ contour map, a 24~$\micron$ source is located near the peak.  There is an H$_{2}$ outflow near the EGO, but they are not physically related.  A $K$-band source coincides with the 4.5~$\micron$ peak and extended toward the SW direction as does the 4.5~$\micron$ emission.

{\it EGO G53.92-0.07 (Figure~\ref{fig:G53.92-0.07})}.  EGO G53.92-0.07 is extended along the EW direction, and is associated with a 24~$\micron$ source.  There is weak H$_{2}$ diffuse emission around the 4.5~$\micron$ peak.  This may be due to residuals from the continuum subtraction rather than being true H$_{2}$ emission.  The $K$-band emission is also extended similar to the 4.5~$\micron$ emission.

{\it EGO G54.11-0.08 (Figure~\ref{fig:G54.11-0.08})}.  The 4.5~$\micron$ emission surrounds a 24~$\micron$ source.  An H$_{2}$ lobe or knot is found around the EGO, but it is unclear if they are related.  There is a $K$-band source located at the 4.5~$\micron$ peak.

{\it EGO G57.61+0.02 (Figure~\ref{fig:G57.61+0.02})}.  The 4.5~$\micron$ is elongated in the NE-SW direction.  There is weak H$_{2}$ emission, but it is not related to the 4.5~$\micron$ emission.  The morphologies of the $K$-band and 4.5~$\micron$ emission are similar.

{\it EGO G58.78+0.64 and G58.78+0.65 (Figure~\ref{fig:G58.78+0.64})}.  In this field, there are two EGOs located at the border of an \ion{H}{2} region.  No 24~$\micron$ sources are associated with these two EGOs.  Three H$_{2}$ knots are found around the EGOs, but it is unclear if they are related.  In the $K$-band image, EGO G58.78+0.64 shows extended emission.

{\it EGO G59.79+0.63 (Figure~\ref{fig:G59.79+0.63})}.  There is a 24~$\micron$ source located at the 4.5~$\micron$ contour peak.  In the continuum-subtracted H$_{2}$ image, an H$_{2}$ lobe is found and the 24~$\micron$ source is likely to be the driving source.  No $K$-band continuum emission is associated with the EGO.

{\it EGO G62.70-0.51 (Figure~\ref{fig:G62.70-0.51})}.  The EGO is extended toward the SE direction from a 24~$\micron$ source located at the 4.5~$\micron$ peak.  There is no H$_{2}$ emission in the continuum-subtracted H$_{2}$ image.  At the 4.5~$\micron$ contour peak, at least three $K$-band sources are located.

\section{H$_{2}$ Outflows Associated with 44~GHz Methanol Masers}\label{appendix}

\begin{figure}
\includegraphics[angle=0,scale=0.4]{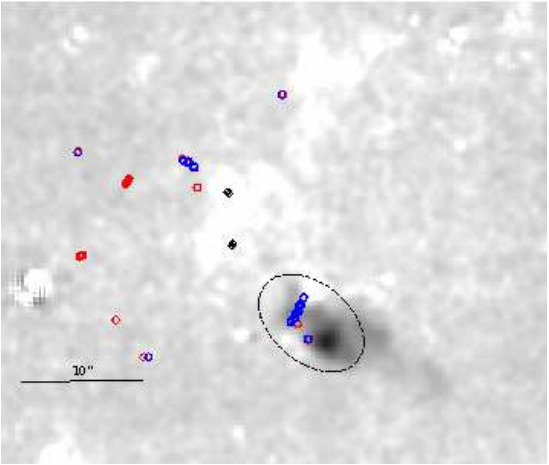}
\caption{Continuum-subtracted H$_{2}$ image for EGO G11.92-0.61, overlaid with the positions of 6.7~GHz and 44~GHz methanol masers.  The dashed black ellipse marks the position of the H$_{2}$ outflow.  The 6.7~GHz (black diamonds) methanol masers mark the positions of the MYSOs \citep{cyg09}.  The red and blue circles are represented the velocities of the 44~GHz methanol masers \citep{cyg09}, red and blue are redshifted and blueshifted to the V$_{LSR}$ of the EGOs, respectively.}
\label{fig:11.92-0.61}
\end{figure}

\begin{figure}
\includegraphics[angle=0,scale=0.4]{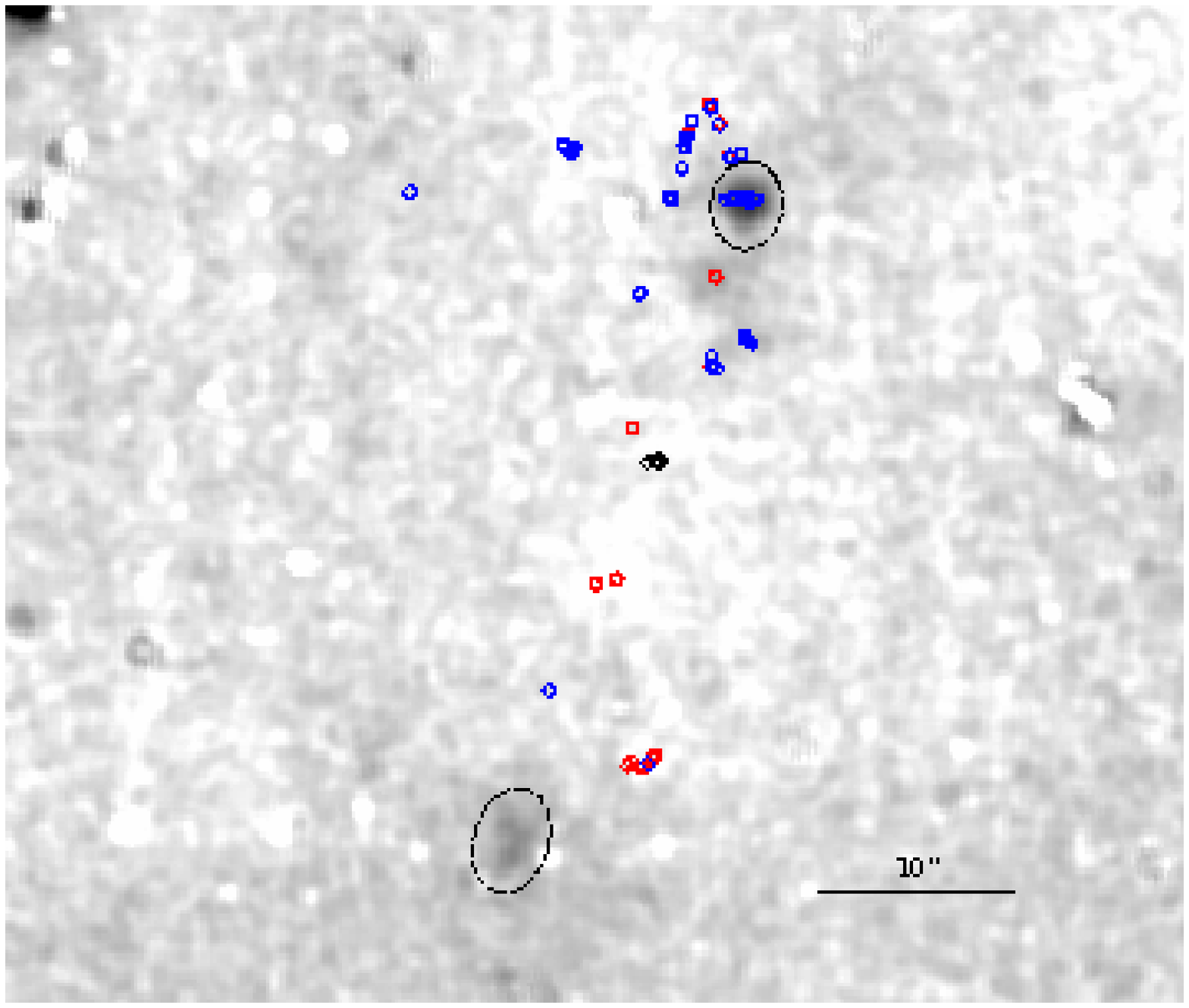}
\caption{Same as Figure~\ref{fig:11.92-0.61}, but for EGO G19.01-0.03.}
\label{fig:19.01-0.03}
\end{figure}

\begin{figure}
\includegraphics[angle=0,scale=0.4]{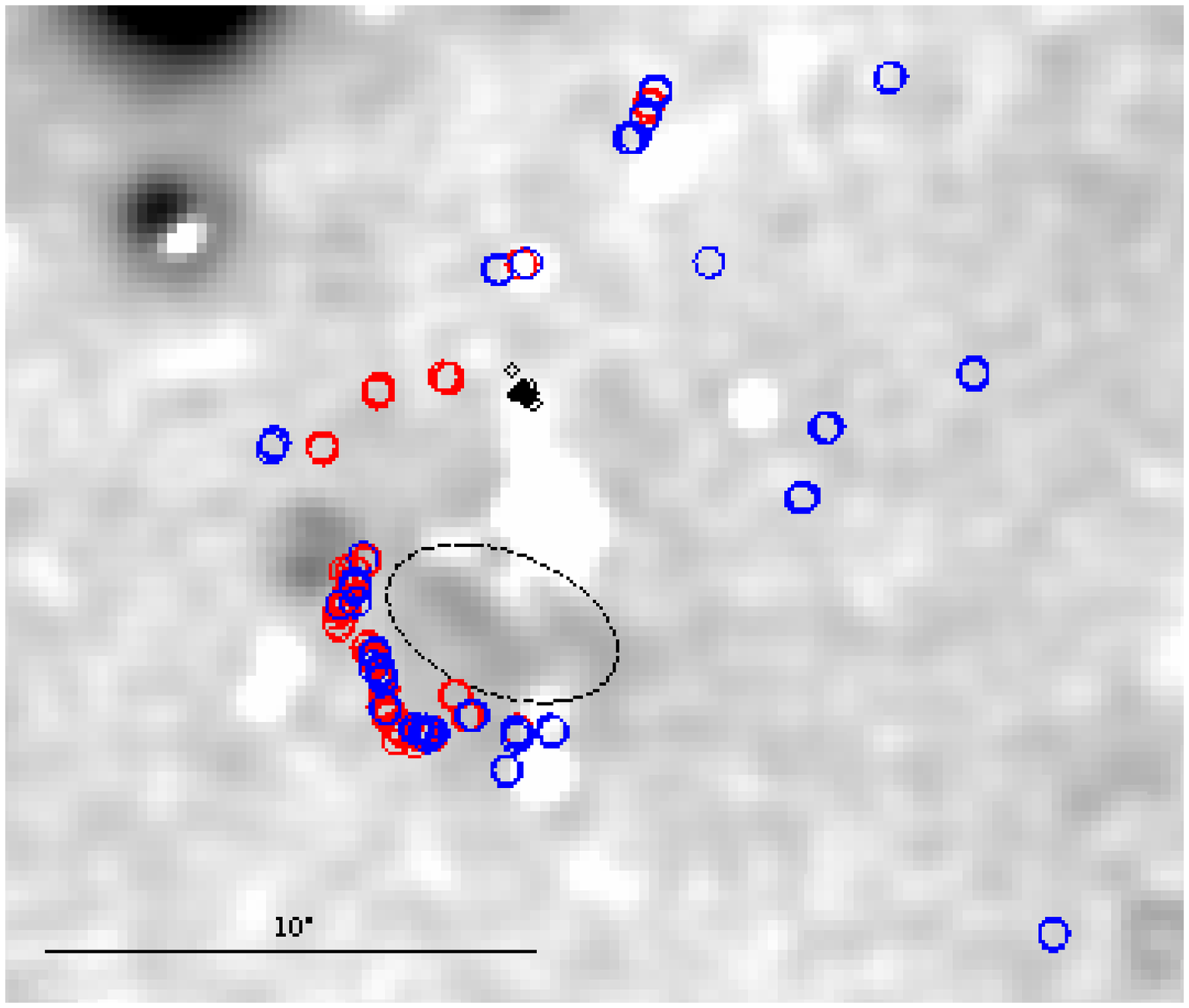}
\caption{Same as Figure~\ref{fig:11.92-0.61}, but for EGO G19.36-0.03.}
\label{fig:19.36-0.03}
\end{figure}

\begin{figure}
\includegraphics[angle=0,scale=0.4]{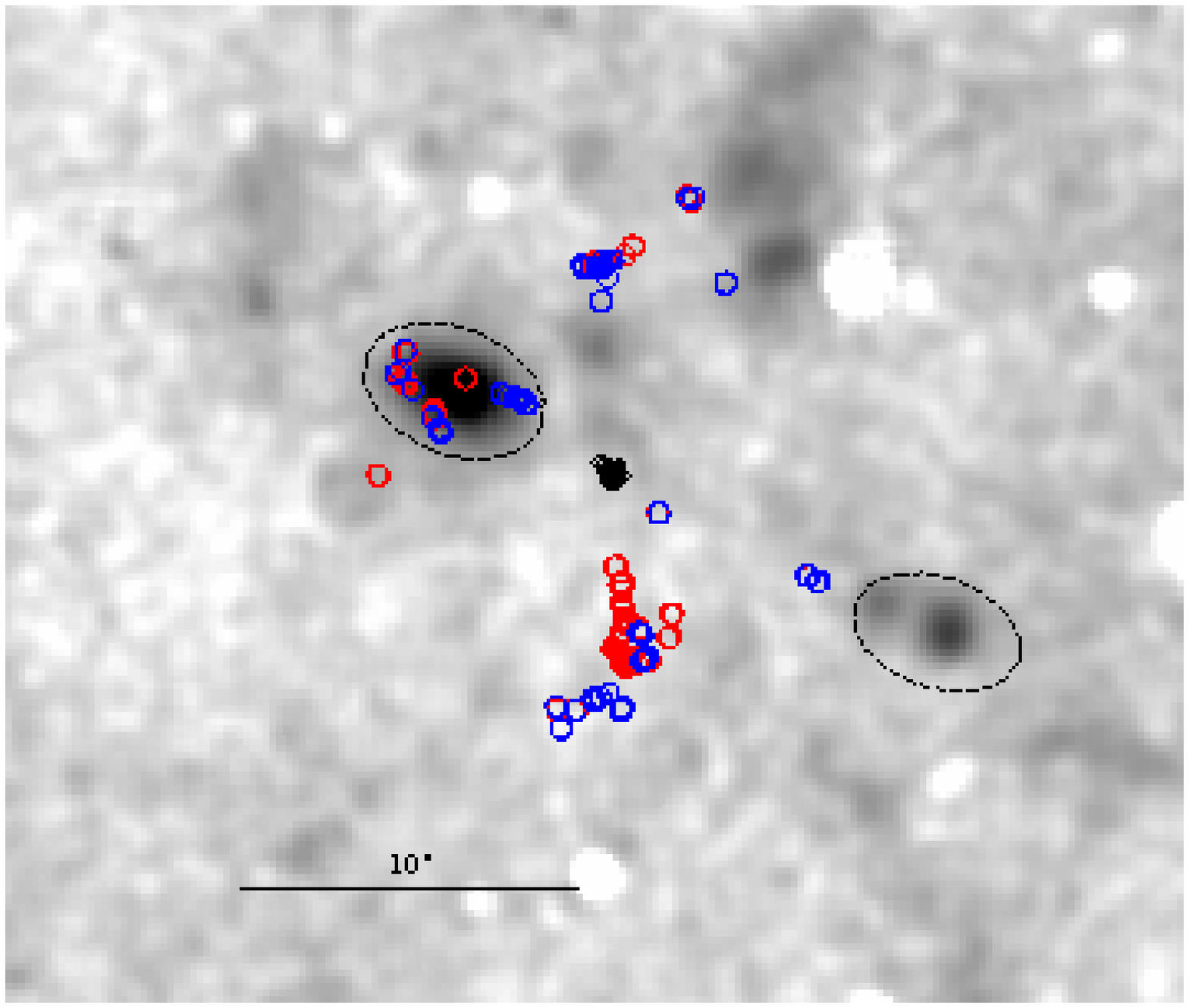}
\caption{Same as Figure~\ref{fig:11.92-0.61}, but for EGO G22.04+0.22.}
\label{fig:22.04+0.22}
\end{figure}

\begin{figure}
\includegraphics[angle=0,scale=0.4]{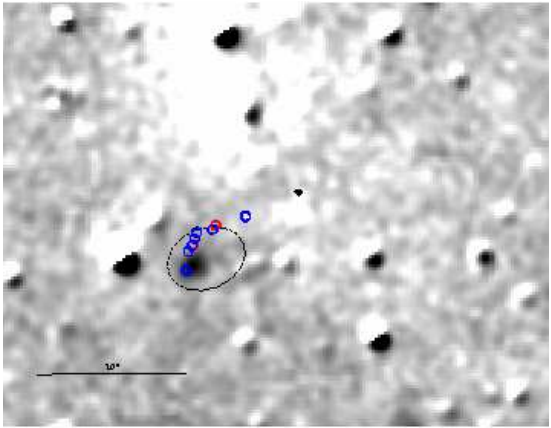}
\caption{Same as Figure~\ref{fig:11.92-0.61}, but for EGO G35.03+0.35.}
\label{fig:35.03+0.35}
\end{figure}

\begin{figure}
\includegraphics[angle=0,scale=0.4]{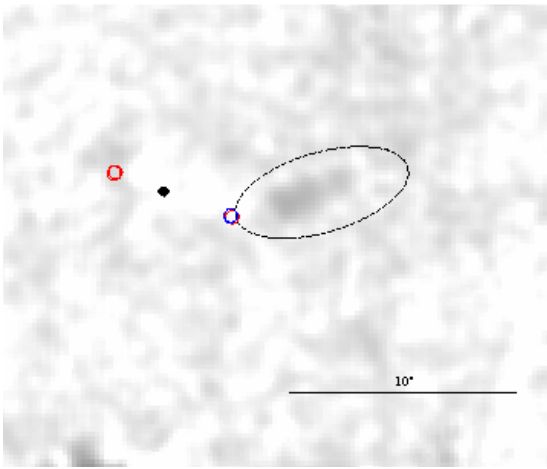}
\caption{Same as Figure~\ref{fig:11.92-0.61}, but for EGO G37.48-0.10.}
\label{fig:37.48-0.10}
\end{figure}

\begin{figure}
\includegraphics[angle=0,scale=0.4]{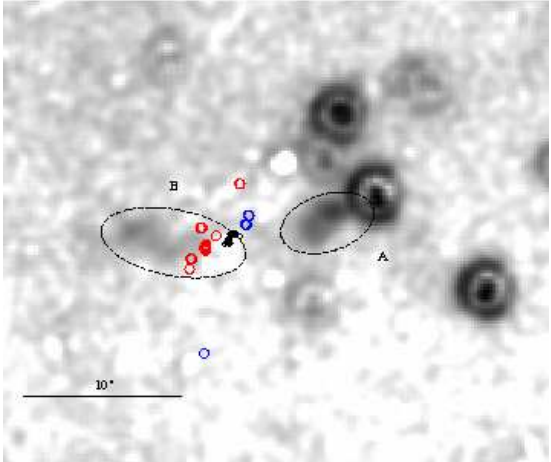}
\caption{Same as Figure~\ref{fig:11.92-0.61}, but for EGO G39.10+0.49.}
\label{fig:39.10+0.49}
\end{figure}

Here, we describe those EGOs with H$_{2}$ detections and 44~GHz methanol maser observations \citep{cyg09}.  In Figure~\ref{fig:11.92-0.61}-\ref{fig:39.10+0.49}, the 44~GHz masers are superposed on the continuum-subtracted H$_{2}$ images.  The red and blue circles are the redshifted and blueshifted masers relative to the V$_{LSR}$, respectively.  The black diamonds represent the positions of the 6.7~GHz methanol masers \citep{cyg09}, tracing the positions of the MYSOs.

{\it EGO G11.92$-$0.61 (Figure~\ref{fig:11.92-0.61})}.  \citet{lee12} found that EGO G11.92-0.61 is associated with an H$_{2}$ outflow lobe.  The majority of the 44~GHz methanol masers around the H$_{2}$ lobe are blueshifted.  The 44~GHz methanol maser group is located upstream of the H$_{2}$ emission.

{\it EGO G19.01$-$0.03 (Figure~\ref{fig:19.01-0.03})}.  For EGO G19.01-0.03, \citet{lee12} identified an H$_{2}$ outflow with two H$_{2}$ knots along the N-S direction.  In the continuum-subtracted H$_{2}$ image, the north H$_{2}$ knot is brighter than the south one.  For the northern H$_{2}$ knot, the 44~GHz methanol masers are distributed both upstream and downstream of the H$_{2}$ emission and the majority of the masers are blueshifted.  In contrast, the 44~GHz methanol masers are only located upstream of the southern H$_{2}$ knot, and most of them are redshifted.

{\it EGO G19.36$-$0.03 (Figure~\ref{fig:19.36-0.03})}.  The EGO is associated with a faint H$_{2}$ lobe.  The 44~GHz methanol masers are distributed downstream of the H$_{2}$ emission, assuming the driving source is located at the position of the 6.7~GHz methanol maser.  In the continuum-subtracted H$_{2}$ image, there are blueshifted and redshifted methanol masers around the H$_{2}$ emission.

{\it EGO G22.04+0.22 (Figure~\ref{fig:22.04+0.22})}.  This EGO has a bipolar H$_{2}$ outflow along the NE-SW direction.  In the continuum-subtracted H$_{2}$ image, the NE H$_{2}$ lobe is brighter than the SW one.  There are blueshifted and redshifted 44~GHz methanol masers located around the NE H$_{2}$ lobe.  In contrast, the SW H$_{2}$ lobe is dominated by blueshifted methanol masers, and the masers are distributed upstream.

{\it EGO G35.03+0.35 (Figure~\ref{fig:35.03+0.35})}.  In the continuum-subtracted H$_{2}$ image, the EGO only shows an H$_{2}$ lobe.  There is a group of 44~GHz methanol masers surrounding the H$_{2}$ emission, and they are mainly blueshifted.  For this EGO, there is no counter-jet for the H$_{2}$ or 44~GHz methanol masers; however \citet{par12} identified a bipolar CO outflow in single dish observations.

{\it EGO G37.48$-$0.10 (Figure~\ref{fig:37.48-0.10})}.  There is an H$_{2}$ lobe located to the west of the 6.7~GHz methanol masers.  Two groups of  methanol masers are distributed along the EW direction.  The eastern group is dominated by redshifted 44~GHz methanol masers.  The western group of the 44~GHz methanol masers is located upstream of the H$_{2}$ lobe, and the majority of the 44~GHz methanol masers are blueshifted.

{\it EGO G39.10+0.49 (Figure~\ref{fig:39.10+0.49})}.  The EGO shows a bipolar H$_{2}$ outflow in the EW direction.  There are redshifted 44~GHz methanol masers distributed upstream of the eastern H$_{2}$ lobe.  However, the brighter western H$_{2}$ lobe is not associated with any 44~GHz methanol masers.

\begin{figure}
\includegraphics[angle=0,scale=0.65]{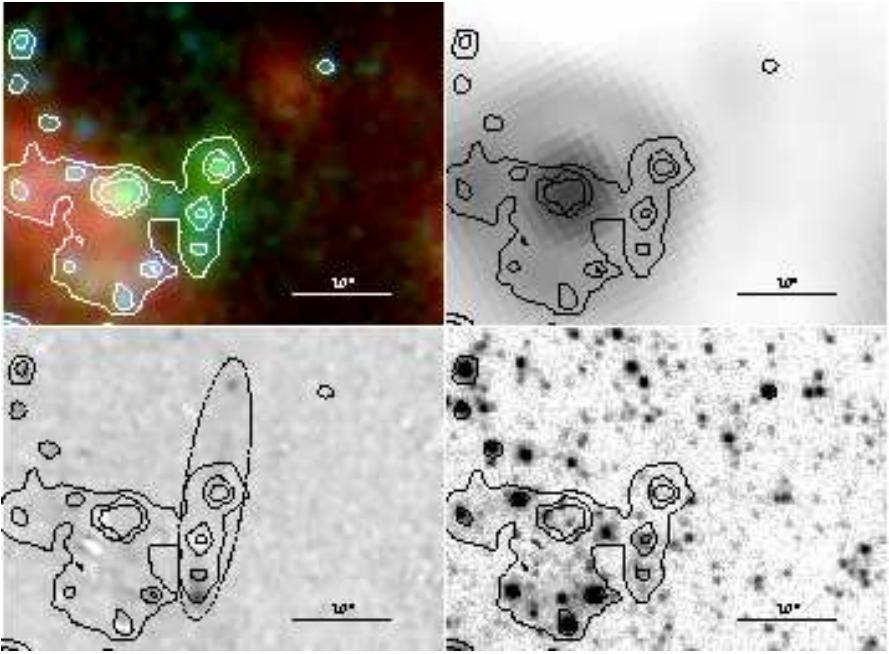}
\caption{Same as Figure~\ref{fig:G11.11-0.11}, but for EGO G49.42+0.33.  The position of EGO G49.42+0.33 is marked by a white dashed ellipse.  The black ellipse marks the H$_{2}$ outflow (MHO 2464).}
\label{fig:G49.42+0.33}
\end{figure}

{\it EGO G49.42+0.33}.  For this EGO, there is no 44~GHz methanol maser detection \citep{cyg09}.  However, there is an H$_{2}$ outflow associated with the EGO (Figure~\ref{fig:G49.42+0.33}).  

\clearpage

\clearpage

\clearpage

\clearpage

\clearpage

\clearpage




\clearpage

\clearpage

\clearpage

\clearpage

\clearpage

\clearpage

\clearpage

\clearpage

\clearpage

\clearpage

\clearpage

\clearpage

\clearpage

\clearpage


\clearpage
\begin{figure}
\includegraphics[angle=0,scale=0.8]{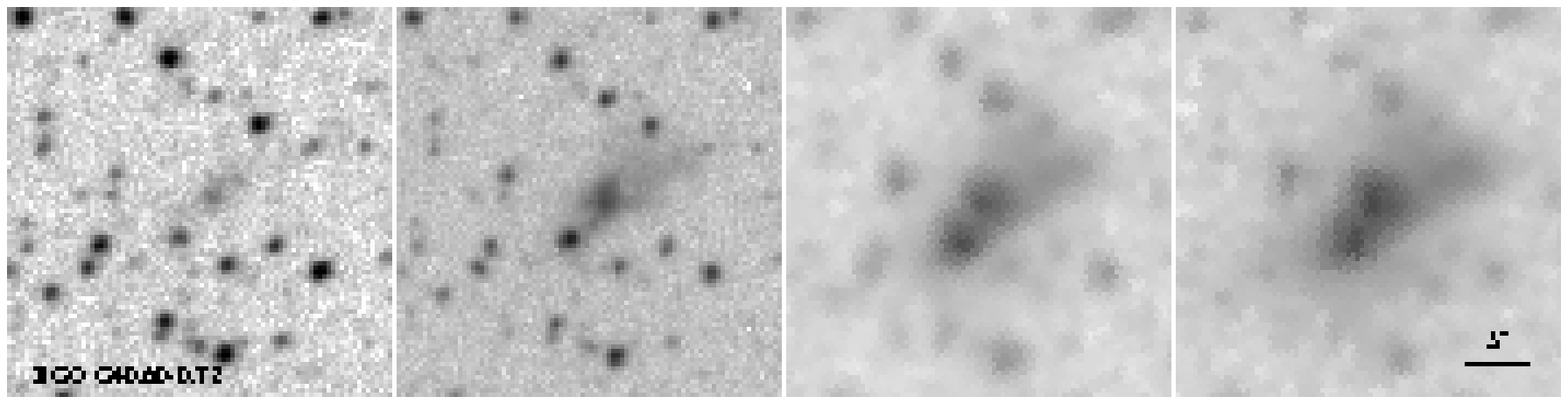}
\includegraphics[angle=0,scale=0.8]{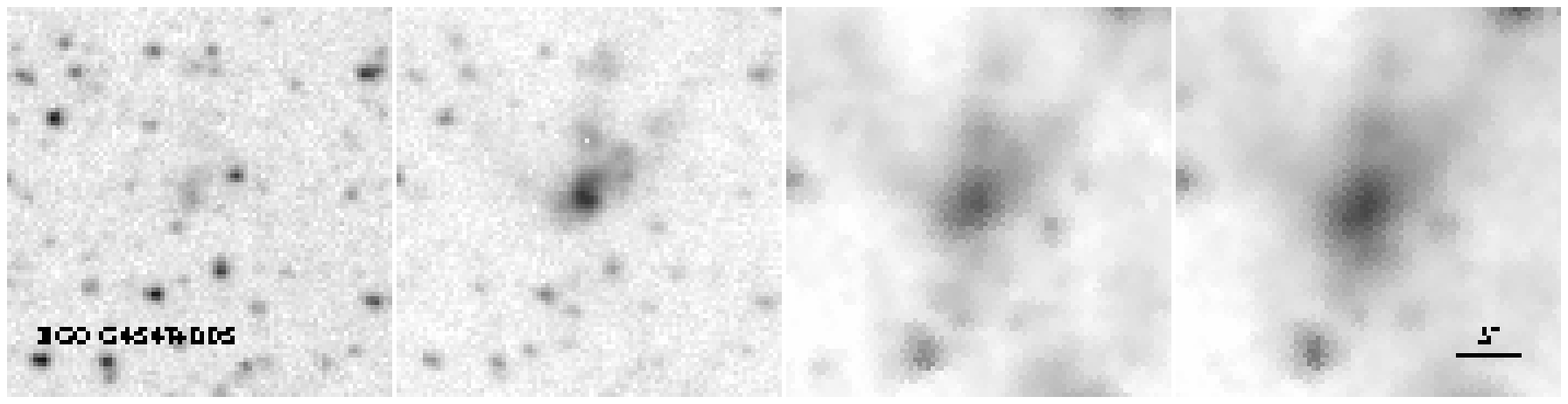}
\includegraphics[angle=0,scale=0.8]{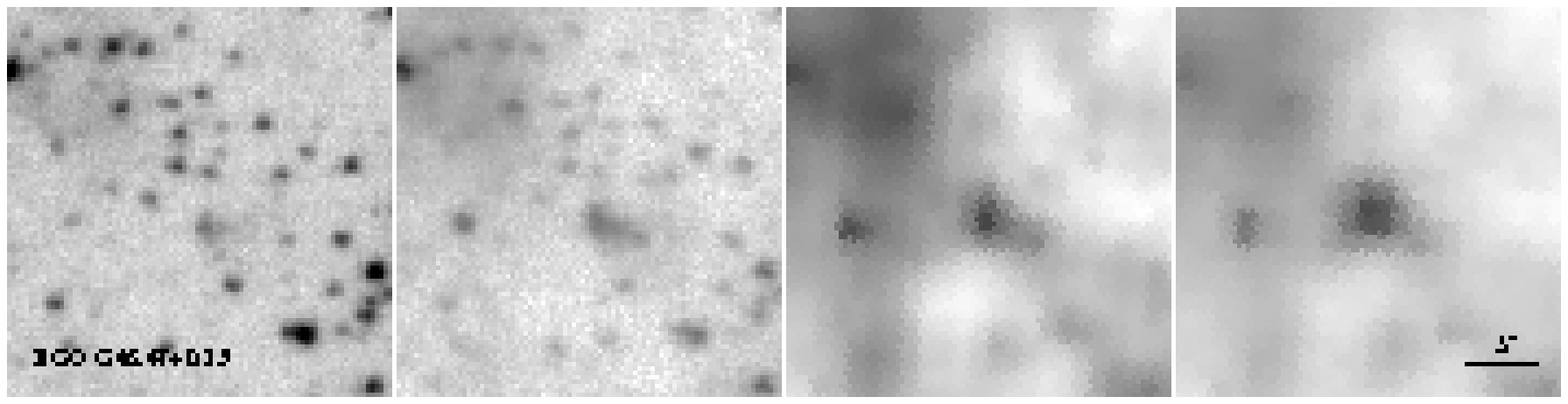}
\includegraphics[angle=0,scale=0.8]{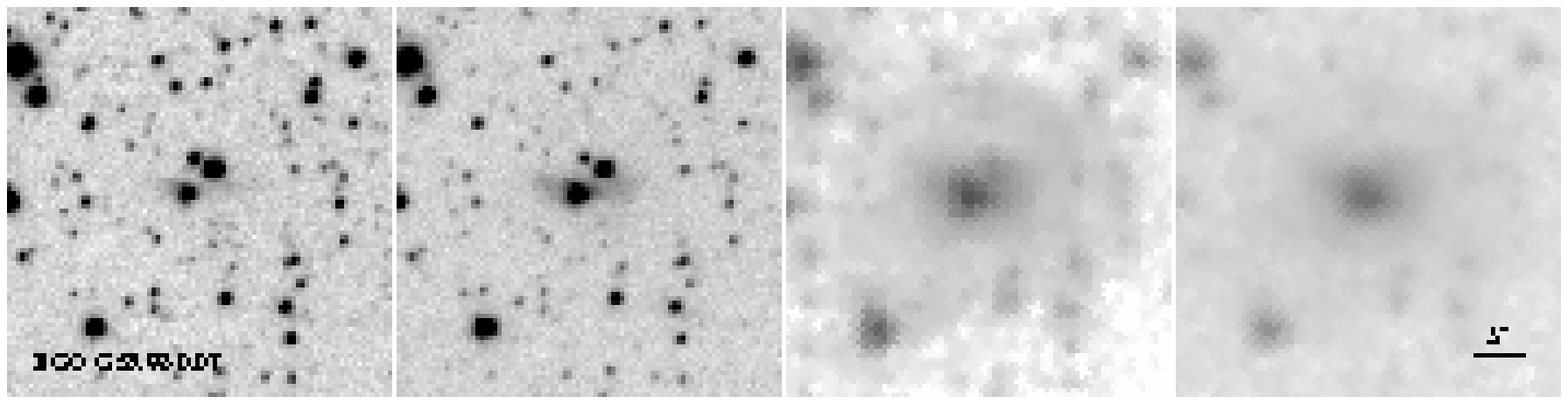}
\includegraphics[angle=0,scale=0.8]{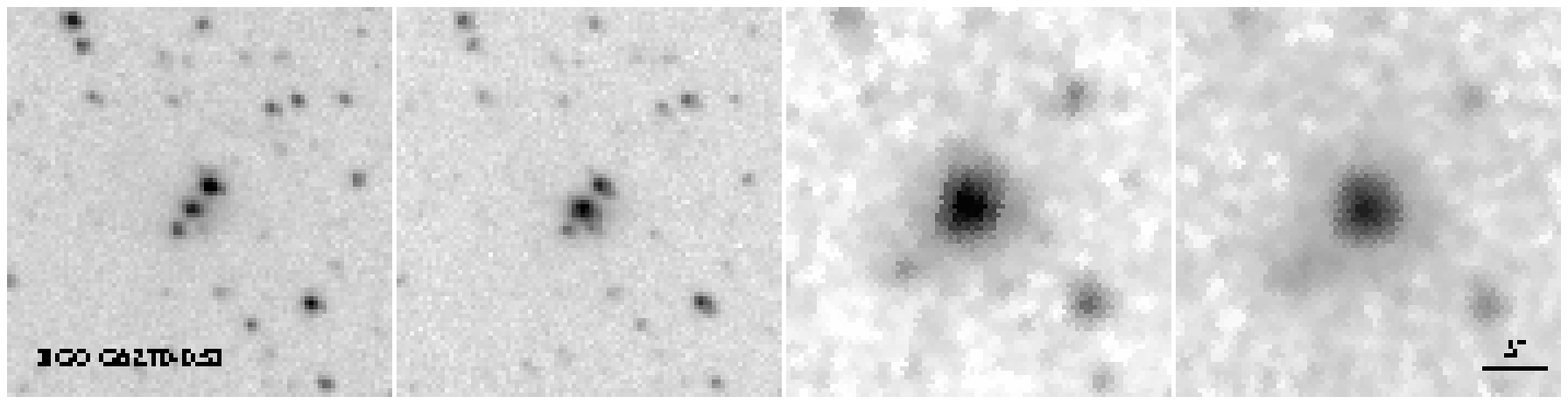}
\centerline{Figure~\ref{fig:continuum} --- Continued.}
\end{figure}

\clearpage


\clearpage

\clearpage

\clearpage

\clearpage

\clearpage

\begin{deluxetable}{lrcrccc}
\tablecaption{H$_{2}$, $K$-, and $H$-band results}
\tablewidth{0pt}
\tabletypesize{\footnotesize}
\tablehead{ & & & \multicolumn{3}{c}{Detection} & \\
\cline{4-6}\noalign{\smallskip}
\colhead{EGO} & \colhead{D.} & \colhead{D. ref.} & \colhead{H$_{2}$} & \colhead{$K$-band} & \colhead{$H$-band} & \colhead{Note}\\
 & \colhead{(kpc)} & & (Y/N) & (Y/N) & (Y/N) & }

\startdata
G11.11$-$0.11    & 3.6 & 1 & Y & Y & N & Two aligned H$_{2}$ knots                                      \\
G12.91$-$0.26    & 3.7 & 1 & Y & Y & Y & A bipolar H$_{2}$ outflow.  HC HII region                      \\
G14.33$-$0.64    & 2.5 & 1 & Y & N & N & Two H$_{2}$ lobes and one H$_{2}$ knot                         \\
G17.96$+$0.08    & 2.3 & 1 & Y?& Y & N & An H$_{2}$ knot                                                \\
G19.36$-$0.03    & 2.5 & 1 & Y?& Y & N &                                                                \\
G21.24$+$0.19    & 2.3 & 1 & Y & Y & Y & A bipolar H$_{2}$ outflow                                      \\
G22.04$+$0.22    & 3.7 & 1 & Y & N & N & A bipolar H$_{2}$ outflow                                      \\
G23.01$-$0.41    & 4.7 & 1 & N & N & N & HC HII region                                                  \\
G23.82$+$0.38    & 4.7 & 1 & Y & N & N & An H$_{2}$ outflow                                             \\
G23.96$-$0.11    & 4.5 & 2 & N & N & N &                                                                \\
G24.00$-$0.10    & 4.5 & 1 & N & N & N &                                                                \\
G24.11$-$0.17    & 4.9 & 1 & N & Y & Y &                                                                \\
G24.11$-$0.18    & ... & - & Y?& Y & N & An H$_{2}$ knot                                                \\
G24.17$-$0.02    & ... & - & N & N & N &                                                                \\
G24.33$+$0.14    & 5.9 & 3 & Y & N & N & An H$_{2}$ outflow and knot, HII                               \\
G24.63$+$0.15    & 3.6 & 1 & Y?& Y & N & A faint H$_{2}$ knot                                           \\
G24.94$+$0.07    & 3.0 & 1 & N & N & N & HC HII region                                                  \\
G25.27$-$0.43    & 3.9 & 2 & N & N & N &                                                                \\
G25.38$-$0.15    & 5.4 & 1 & N & N & N &                                                                \\
G27.97$-$0.47    & ... & - & Y & Y & N & A bipolar H$_{2}$ outflow                                      \\
G28.28$-$0.36    & 3.3 & 1 & N & N & N & HC HII region                                                  \\
G29.84$-$0.47    & ... & - & N & N & N &                                                                \\
G29.89$-$0.77    & 5.0 & 1 & N & N & N &                                                                \\
G29.91$-$0.81    & 5.0 & 1 & N & N & N &                                                                \\
G29.96$-$0.79    & 5.1 & 1 & Y & Y & Y & An H$_{2}$ outflow                                             \\
G34.26$+$0.15    & 3.8 & 1 & Y?& Y & N & Several H$_{2}$ knots. Two HC~HII regions                      \\
G34.28$+$0.18    & 3.8 & 1 & N & Y & N &                                                                \\
G34.39$+$0.22    & 3.6 & 3 & Y & Y & N & An H$_{2}$ lobe                                                \\
G34.41$+$0.24    & 3.6 & 3 & N & N & N &                                                                \\
G35.03$+$0.35    & 3.5 & 1 & Y & Y & N & An H$_{2}$ lobe. UC HII and HC HII regions                     \\
G37.48$-$0.10    & 3.8 & 1 & Y & Y & N & An H$_{2}$ lobe                                                \\
G37.55$+$0.20    & ... & - & N & N & N & HC HII region                                                  \\
G39.10$+$0.49    & 1.7 & 1 & Y & Y & N & A bipolar H$_{2}$ outflow                                      \\
G39.39$-$0.14    & 4.5 & 1 & N & Y & N & HII region                                                     \\
G40.28$-$0.22    & 5.4 & 1 & Y & Y & Y & An H$_{2}$ lobe                                                \\
G40.28$-$0.27    & 5.2 & 1 & Y?& Y & N & An H$_{2}$ knot                                                \\
G40.60$-$0.72    & 4.6 & 1 & N & Y & Y &                                                                \\
G43.04$-$0.45(a) & 4.2 & 1 & N & N & N &                                                                \\
G43.04$-$0.45(b) & 4.2 & 1 & N & N & N &                                                                \\
G44.01$-$0.03    & 5.3 & 1 & N & N & N &                                                                \\
G45.47$+$0.05    & 5.0 & 1 & N & Y & Y & HII region                                                     \\
G45.47$+$0.13    & 5.2 & 1 & N & Y & Y &                                                                \\
G45.50$+$0.12    & 5.3 & 1 & N & N & N &                                                                \\
G45.80$-$0.36    & 4.6 & 3 & N & Y & N &                                                                \\
G48.66$-$0.30    & 2.7 & 1 & Y & N & N & A bipolar H$_{2}$ outflow                                      \\
G49.07$-$0.33    & 5.0 & 1 & N & Y & N &                                                                \\
G49.27$-$0.32    & 5.0 & 1 & N & N & N &                                                                \\
G49.27$-$0.34    & 5.0 & 1 & N & Y & N & UCHII region                                                   \\
G49.42$+$0.33    &12.3 & 1 & Y & Y & N & A bipolar H$_{2}$ outflow                                      \\
G49.91$+$0.37    & 0.8 & 3 & N & N & N &                                                                \\
G50.36$-$0.42    & 3.2 & 3 & N & Y & N &                                                                \\
G53.92$-$0.07    & 4.6 & 1 & Y?& Y & Y & Extended H$_{2}$ emission                                      \\
G54.11$-$0.08    & 4.3 & 1 & Y?& Y & N & An H$_{2}$ lobe or knot                                        \\
G56.13$+$0.22    & ... & - & N & N & N &                                                                \\
G57.61$+$0.02    & 4.5 & 3 & Y?& Y & N & H$_{2}$ extended emission                                      \\
G58.78$+$0.64    & 5.0 & 1 & Y?& Y & N & Three H$_{2}$ knots                                            \\
G58.78$+$0.65    & ... & - & Y?& N & N & Three H$_{2}$ knots                                            \\
G58.79$+$0.63    & ... & - & N & N & N &                                                                \\
G59.79$+$0.63    & 5.0 & 1 & Y & N & N & An H$_{2}$ lobe                                                \\
G62.70$-$0.51    & ... & - & N & Y & Y &                                                                \\

\enddata
\tablerefs{(1) \citet{che09}; (2) \citet{cyg09}; (3) \citet{he12}}
\label{tab:h2}
\end{deluxetable}

\clearpage

\begin{deluxetable}{lrll}
\tablecaption{New MHO numbers in this paper}
\tablewidth{0pt}
\tabletypesize{\footnotesize}
\tablehead{\colhead{MHO number} & \colhead{RA (J2000)} & \colhead{DEC (J2000)} & \colhead{Note}}

\startdata
2304   & 18 10 28.3 & -19 22 31 & An H$_{2}$ lobe from EGO G11.11-0.11                                  \\
2305   & 18 14 39.5 & -17 52 00 & A bipolar H$_{2}$ outflow along the EW direction from EGO G12.91-0.26 \\
2306   & 18 18 54.4 & -16 47 46 & Two H$_{2}$ lobe around EGO G14.33-0.64                               \\
3258   & 18 35 08.1 & -07 35 04 & A bipolar H$_{2}$ outflow from EGO G24.33+0.14                        \\
2458   & 18 48 50.0 & -03 00 21 & An H$_{2}$ outflow from EGO G29.96-0.79                               \\
2459   & 18 53 19.0 & +01 24 08 & An H$_{2}$ lobe from EGO G34.39+0.22                                  \\
2460   & 18 54 00.5 & +02 01 18 & An H$_{2}$ lobe from EGO G35.03+0.35                                  \\
2461   & 19 00 07.0 & +03 59 53 & An H$_{2}$ lobe from EGO G37.48-0.10                                  \\
2462   & 19 00 58.1 & +05 42 44 & A bipolar H$_{2}$ outflow from EGO G39.10+0.49                        \\
2463   & 19 05 41.3 & +06 26 13 & An H$_{2}$ lobe from EGO G40.28-0.22                                  \\
2464   & 19 20 59.1 & +14 46 53 & A bipolar H$_{2}$ outflow from EGO G49.42+0.33                        \\
2465   & 19 21 48.0 & +13 49 21 & A bipolar H$_{2}$ outflow from EGO G48.66-0.30                        \\
2624   & 19 25 31.5 & +15 15 50 & A bipolar H$_{2}$ outflow near EGO G50.36-0.42                        \\
2625   & 19 41 03.1 & +24 01 15 & An H$_{2}$ lobe from EGO G59.79+0.63                                  \\
\enddata
\label{tab:mho}
\end{deluxetable}

\begin{deluxetable}{ll}
\tablecaption{EGOs with known radio continuum detections}
\tablewidth{0pt}
\tablehead{\colhead{EGO} & \colhead{reference}}

\startdata

\multicolumn{2}{c}{UC~HII}  \\

G12.20$-$0.03 & \citet{urq09} \\
G12.42$+$0.50 & \citet{urq09} \\
G24.33$+$0.14 & \citet{bat10} \\
G35.03$+$0.35 & \citet{cyg11} \\
G39.39$-$0.14 & \citet{hof11} \\
G45.47$+$0.05 & \citet{urq09} \\
G49.27$-$0.34 & \citet{cyg11} \\
\\
\multicolumn{2}{c}{HC~HII}  \\

G11.92$-$0.61 & \citet{cyg11} \\
G12.91$-$0.26 & \citet{van05} \\
G23.01$-$0.41 & \citet{san10} \\
G24.94$+$0.07 & \citet{cyg11} \\
G28.28$-$0.36 & \citet{cyg11} \\
G28.83$-$0.25 & \citet{cyg11} \\
G34.26$+$0.15 & \citet{gau94,sew11} \\
G35.03$+$0.35 & \citet{cyg11} \\
G35.20$-$0.74 & \citet{gib03,beu05} \\
G37.55$+$0.20 & \citet{san11} \\
\\
\multicolumn{2}{c}{Thermal jet}  \\

G16.59$-$0.05 & \citet{zap06} \\
G19.88$-$0.53 & \citet{zap06} \\

\enddata
\label{tab:radio}
\end{deluxetable}

\end{document}